\documentclass[a4paper,11pt]{article}
\usepackage{jcappub,aas_macros} 
\usepackage{lineno}
 \usepackage{stmaryrd}

\usepackage{mathtools}
\newcommand{\bq}{\boldsymbol q}

\newcommand{\hq}{\hat{q}}
\newcommand{\hk}{\hat{k}}
\newcommand{\hn}{\hat{n}}

\newcommand{\bx}{\boldsymbol x}
\newcommand{\bv}{\boldsymbol v}
\newcommand{\bu}{\boldsymbol u}
\newcommand{\bk}{\textbf{k}}

\newcommand{\bs}{\textbf{s}}

\usepackage[normalem]{ulem}

\newcommand{\avg}[1]{\left\langle #1 \right\rangle}

\newcommand{\edit}[1]{{{{#1}}}}

\title{Effective Theories of Redshift-Space Galaxy Peculiar Velocities}

\author[a,b,c]{Shi-Fan Chen,}
\author[d]{Cullan Howlett,}
\author[d]{Yan Lai,}
\author[e]{Fei Qin}
\affiliation[a]{School of Natural Sciences, Institute for Advanced Study, 1 Einstein Drive, Princeton, NJ
08540}
\affiliation[b]{Physics Department, Columbia University, Pupin Hall, 538 West 120th Street 704, MC 5255, New York, NY 10027}
\affiliation[c]{Einstein Fellow, NASA Hubble Fellowship Program}
\affiliation[d]{School of Mathematics and Physics, The University of Queensland, Brisbane, QLD 4072, Australia}
\affiliation[e]{Aix-Marseille University, CNRS/IN2P3, CPPM, Marseille 13288, France}
\emailAdd{sfschen@ias.edu}

\abstract{We present predictions for redshift-space peculiar velocity statistics in the Lagrangian and Eulerian formulations of the effective field theory (EFT) of large-scale structure. We compute 2-point pairwise velocity statistics up to the second moment at next-to-leading (1-loop) order, showing that they can be modeled together with redshift-space galaxy densities with a consistent set of EFT coefficients. We show that peculiar velocity statistics have a distinct dependence on long-wavelength bulk flows that necessitates a variation on the usual infrared (IR) resummation procedure used to model baryon acoustic oscillations (BAO) in galaxy clustering. This can be implemented recursively in powers of the velocity in both the Lagrangian and Eulerian frameworks. We validate our analytic calculations against fully nonlinear N-body simulations, demonstrating that they can be used to recover the growth rate at better than percent level precision, well beyond the statistical requirements of upcoming peculiar velocity surveys and measurements of the kinetic Sunyaev-Zeldovich  (kSZ) effect. As part of this work, we release \href{https://github.com/sfschen/velocisaurus}{\texttt{velocisaurus}}, a fast \texttt{Python} code for computing EFT predictions of peculiar velocity statistics. }

\begin{document}
\maketitle
\flushbottom

\section{Introduction}
\label{sec:intro}

On large, cosmological scales dominated by gravity, general relativity and the standard model of cosmology --- along with its extensions --- make precise and constraining predictions about the growth of structure over cosmic time. A particularly interesting prediction for theories of structure formation are the peculiar velocities of galaxies. Due to the equivalence principle, these directly probe the gravitational potential on large scales, making them an ``unbiased'' tracer of cosmic structure. The premier cosmological measurements using galaxy peculiar velocities comes from spectroscopic surveys such as the Dark Energy Spectroscopic Instrument (DESI) \cite{DR1_fullshape}, where they manifest as perturbations of the measured redshifts of galaxies. This introduces an anisotropy in galaxy clustering known as redshift-space distortions (RSD). Particularly at higher redshifts, where we can probe large survey volumes, RSD from current and upcoming galaxy surveys \cite{SpecS5} promise to make measurements of the rate (i.e. through the so-called growth rate $f\sigma_8$) and scale-dependence of structure formation at the percent level or below. At low redshifts in the nearby universe, on the other hand, the relatively small number of Fourier modes available on large-scales makes precise measurements of the growth rate through RSD more challenging. 

For the nearby Universe a promising alternative is to measure peculiar velocities directly by calibrating their LOS distances independently of redshift. This can be done, for example, by using the fundamental plane relation \citep{Djorgovski1987,Dressler1987}, the Tully-Fisher relation \citep{Tully1977} or through Type Ia supernovae observations \citep{Phillips1993}. While still limited by the small volumes available at low redshifts, peculiar velocities have the advantage that their signal-to-noise at large scales goes as $k^{-2} n_{v} P(k)$ rather than $n_{\delta} P(k)$ as in case of galaxies \citep{Howlett19,Giani2023}, where $n_{v,\delta}$ are the constant shot noises for velocities and densities and $P(k)$ is the linear power spectrum. On large scales, peculiar velocity measurements are therefore dominated by cosmic variance and faithfully sample large-scale modes. Furthermore, a combined analysis of redshifts and velocities in the local Universe allows us to overcome the effects of sample variance  \citep{Burkey2004,Koda2014,Howlett2017a}, in the same way as other multi-tracer techniques \citep{Mcdonald2009b}. Motivated by this, current state-of-the-art data \citep{Springob2014,Tully2016,Hong2019,Qin2021,Howlett2022,Tully2023} has already been used to place constraints on the growth rate of structure using combined clustering analyses of redshifts and peculiar velocities \cite{Johnson2014,Howlett2017c,Adams2017,Qin2019b,Adams2020,Lai23,Turner2023,Qin2023,Lyall2024,Qin2025}. Upcoming, even larger peculiar velocity surveys from DESI \cite{DESI_PVS}, 4MOST \citep{Taylor2023}, ZTF \citep{Carreres2023}, WALLABY \cite{Koribalski2020} and the Rubin Observatory \citep{Howlett2017b, Rosselli2025} are expected to yield even more significant advances in the constraining power of PV measurements, improving by factors of $2-3$ on what can be achieved using only RSD.

In the past decade, significant advances have been made towards analytic predictions of spectroscopic galaxy clustering and RSD, particularly using techniques from effective field theory (EFT) \cite{Ivanov20,DAmico20,Chen22}. These methods allow for the consistent prediction of large-scale galaxy clustering statistics from first-principles, accounting for the backreaction of small-scale astrophysical unknowns and nonlinear dynamics by systematically enumerating their contributions parametrized by free coefficients in the EFT \cite{McDonald09,Senatore12,Porto14,Vlah15,Perko16,Desjacques18,Chen20}. The predictions of these so-called EFTs of large-scale structure (LSS) have become a standard tool for the cosmological analysis of galaxy clustering, and have been shown in many independent studies to have theoretical errors well below the statistical requirements of current and future galaxy surveys on cosmological scales of interest \cite{Nishimichi20,Ivanov23,DAmico24,Maus25b}. On the other hand, while peculiar velocity surveys operate at low redshifts where gravitational and astrophysical nonlinearities are the most pronounced, there has so far been limited theoretical attention to predicting their observables (but see e.g. refs.~\cite{Okumura14,Sugiyama16,Howlett19,Dam21}).

The goal of this paper is to update the modeling of peculiar velocity statistics using the EFT formalism developed for predictions of galaxy clustering. Indeed, redshift-space galaxy clustering derives its anisotropy from the very same LOS peculiar velocities measured by peculiar velocity surveys, leading to strict mathematical relations between the two types of observables we can exploit \cite{Sugiyama16}---we describe these in more detail in Section~\ref{sec:pv_overview} where we also introduce the pairwise velocity statistics we will study in this work. The rest of the paper is structured as follows: In Section~\ref{sec:pt_overview}, we introduce the Eulerian and Lagrangian formulations of the EFT, and give a quick review of infrared (IR) resummation \cite{Baldauf15,Senatore15,Vlah15,Vlah16b,Blas16,Chen24} --- a procedure used to tame large contributions due to long-wavelength bulk flows and velocities --- before describing the extension of these formalisms to peculiar velocities in Section~\ref{sec:pv_pt}. The velocity IR resummation procedure turns out to be surprisingly different than that in galaxy densities, and we explore it in the context of a toy model in Section~\ref{sec:pv_zel}. Finally, we compare our analytic predictions to the peculiar velocities of dark matter halos in fully nonlinear N-body simulations in Section~\ref{sec:nbody}, before concluding in Section~\ref{sec:conclusions}.

\section{Peculiar Velocities and Redshift-Space Distortions}
\label{sec:pv_overview}

\subsection{Redshift-Space Statistics}

The positions of galaxies in cosmological surveys are typically inferred from their measured redshifts $z$. Since these redshifts contain contributions from galaxy peculiar velocities $\bv_p$, the inferred positions are shifted from the ``true'' coordinates $\bx$ of each galaxy to redshift-space coordinates $\bs = \bx + \bu(\bx)$, where we have re-expressed the line-of-sight peculiar velocity in units of distance using $\bu = (\hn \cdot \bv_p) \hn / \mathcal{H}$, and where $\mathcal{H}$ is the conformal Hubble parameter. Since the mapping from real to redshift space conserves the number of galaxies we can relate the redshift-space overdensity $\delta_{g,s}(\bs)$ of galaxies to the real-space overdensity $\delta_g$ via \cite{Scoccimarro04}
\begin{equation}
    1 + \delta_{g,s}(\bs) = \int d^3\bx \ (1 + \delta_g(\bx)) \ \delta_D(\bs-\bx-\bu(\bx))
    \label{eqn:number_conservation}
\end{equation}
where $\delta_D$ is the Dirac delta function, and equivalently in Fourier space
\begin{equation}
    (2\pi)^3 \delta_D(\bk) + \delta_{g,s}(\bk) = \int d^3 \bx\ e^{-i\bk\cdot\bx} e^{-i\bk\cdot\bu(\bx)} \ (1 + \delta_g(\bx)).
\end{equation}
This mapping induces an additional anisotropy to the observed clustering of galaxies known as redshift-space distortions. The extension of these expressions to redshift-space peculiar velocity fields is straightforward --- for example, the line-of-sight momentum density in redshift-space $\rho_{g,s}$ can be obtained by substituting the real-space line-of-sight momentum field $\rho_{g}=(1 + \delta_g) \bu$ in place of the \edit{mean-density normalized} real-space density field $n_g = 1 + \delta_g$ in Equation~\ref{eqn:number_conservation}.

In this work our primary focus will be on developing predictions for 2-point statistics of peculiar velocities in redshift space. The starting point for these predictions will be the galaxy power spectrum, or Fourier-space 2-point function, in redshift space. Combining the above equations with translation invariance, the power spectrum can be written as \cite{Scoccimarro04,Vlah19}
\begin{equation}
    P_s(\bk ; \lambda) = \int d^3 \bx\ e^{i\bk\cdot\bx} \avg{e^{i\bk\cdot(\lambda \Delta\bu)}\ (1 + \delta_{g}(\bx_1))(1+\delta_{g}(\bx_2))}_{\bx=\bx_2-\bx_1}
    \label{equation:power_dpectrum}
\end{equation}
where we have defined the pairwise velocity $\Delta \bu = \bu_2 - \bu_1$.

Let us note two useful properties of Equation~\ref{equation:power_dpectrum}. Firstly, we have inserted an auxiliary $\lambda$ into the exponent to use as a counting parameter for peculiar velocities. The redshift-space power spectrum is given by Equation~\ref{equation:power_dpectrum} evaluated with $\lambda = 1$, while the real-space power spectrum without peculiar velocities is recovered when $\lambda = 0$. Secondly, in the flat sky limit it is invariant under Galilean transformations by a constant velocity $\bv_p \rightarrow \bv_p + \bv_{\rm const}$, such that the measured power spectrum is not affected by large-scale bulk flows. The flat-sky qualifier is necessary because the line-of-sight direction $\hn$ is not in general constant. In the rest of this work we will not consider these wide angle effects, focusing rather on extending the reach of perturbative methods to smaller scales,  using effective-theory methods, where these effects are small. However, we refer the interested reader to e.g. refs.~\cite{Castorina19,Pantiri24} for excellent treatments of wide-angle and other large-scale effects from general relativity.

\subsection{Line-of-Sight Velocity Statistics}

We are interested in two point statistics of powers of the galaxy peculiar velocity $\bu$ along the line-of-sight. Specifically, we are interested in spectra of the form \cite{Gorski89,Groth89,Park2000}
\begin{equation}
    P^s_{LL'}(\bk) =  \int d^3 \bs\ e^{i\bk\cdot\bs} \avg{(1 + \delta_{g,s}(\bs_1))(1+\delta_{g,s}(\bs_2))\ \bu_{s,\hn}^L(\bs_1) \bu_{s,\hn}^{L'}(\bs_2) }_{\bs=\bs_1-\bs_2}
\end{equation}
which can be expressed in terms of odd or even multipoles $P_{LL'}^\ell(k)$ depending on the parity $L + L'$. The spectra above can be re-expressed as a real-space integral using the pairwise velocity $\Delta \bu$ as \cite{Ferreira99,Okumura14}
\begin{equation}
    P^s_{LL'}(\bk) = \int d^3 \bx\ e^{i\bk\cdot\bx} \avg{e^{i\bk\cdot\Delta\bu}\ (1 + \delta_{g}(\bx_1))(1+\delta_{g}(\bx_2))\ \bu_{\hn}^L(\bx_1) \bu_{\hn}^{L'}(\bx_2) }_{\bx=\bx_2-\bx_1}.
\end{equation}
The redshift-space galaxy power spectrum is given in terms of the above by $L = L' = 0$, while for example the density-momentum cross spectrum is given by $L, L' = 1, 0$ and the momentum autospectrum by $L = L' = 1$.

A particularly nice combination of the velocity spectra above are density-weighted \textit{pairwise} velocity spectra. These are defined as \cite{Juszkiewicz89,Sugiyama16}
\begin{equation}
    \tilde{\Xi}_s^{(n)}(\bk) = \int d^3 \bs\ e^{i\bk\cdot\bs} \avg{(1 + \delta_{g,s}(\bs_1))(1+\delta_{g,s}(\bs_2))\ \Delta \bu_{\hn}^n }_{\bs=\bs_1-\bs_2}
    \label{eqn:pairwise_vels}
\end{equation}
\edit{where the tildes denote that they are Fourier-space quantities}. These can again be expressed in terms of their multipole moments $\tilde{\Xi}^{(n)}_{s,\ell}$, and are given by linear combinations of $P_{LL'}$ where $n = L + L'$. In this work we will be particularly interested in the first and second moments of the pairwise velocity 
\begin{equation}
v_s(\bk) = \tilde{\Xi}^{(1)}_s = 2 P^s_{01}, \quad 
\sigma^2_s(\bk) = \tilde{\Xi}^{(2)}_s = 2P^s_{02} - 2 P_{11},
\end{equation} 
which specify the mean infall velocity between galaxies and its dispersion. We will also denote their real-space counterparts by dropping the subscript ``s.'' As in the case of the redshift-space power spectrum, since these spectra only depend upon pairwise velocities they are Galilean invariant in the flat sky limit. 

Given a model of the redshift-space power spectrum, the redshift-space pairwise-velocity spectra in Equation~\ref{eqn:pairwise_vels} are uniquely specified. In fact, the redshift-space power spectrum is their generating function, with the pairwise spectra given by derivatives \cite{Sugiyama16}
\begin{equation}
    \tilde{\Xi}_s^{(n)}(\bk) = \frac{1}{(ik\mu)^n}\left( \frac{d^n P_s(\bk;\lambda)}{d \lambda^n} \right)_{\lambda=1}.
    \label{eqn:generating_function}
\end{equation}
A subtlety, as we will see, is that many effective theory parameters which enter the redshift-space power spectrum in degenerate ways have distinct $\lambda$ dependence, which will introduce new free parameters into the final model. Similarly, real-space pairwise velocity spectra $\tilde{\Xi}^{(n)}$ can be obtained by setting $\lambda=0$ in Equation~\ref{eqn:generating_function} \cite{Scoccimarro04}. Conversely, the redshift-power spectrum can be built order-by-order using real-space pairwise velocity spectra \cite{Seljak11,Vlah19}
\begin{equation}
    P_{s}(\bk;\lambda) = \sum_{n=0}^\infty \frac{(ik\mu)^n \lambda^n }{n!} \tilde{\Xi}^{(n)}(\bk).
    \label{eqn:mome}
\end{equation}
This is known in the literature as the distribution function or moment expansion approach to redshift-space distortions. Indeed, it is generically possible to construct the redshift-space velocity statistics as a Taylor series with coefficients proportional to real-space pairwise velocity spectra, an approach that will be especially useful for our Eulerian perturbation theory predictions in the sections below \cite{Okumura14}.

\section{Redshift-Space Power Spectra: Effective Theories and IR Resummation}
\label{sec:pt_overview}

Let us now review the perturbative calculation of the galaxy redshift-space power spectrum in the effective theory of large-scale structure. Since the redshift-space power spectrum is the generating function of pairwise velocity statistics, these calculations will be essential tools with which we can compute redshift-space peculiar velocity statistics. Specifically, we will discuss these calculations in the formalisms of Lagrangian and Eulerian perturbation theory, reviewing in particular effective-theory contributions like counterterms and stochastic terms and the infrared (IR) resummation of baryon acoustic oscillations (BAO). Throughout this paper we will work at 1-loop order in perturbation theory, which is sufficient for next-to-leading-order predictions of the observables we are interested in.

\subsection{Lagrangian Perturbation Theory}
\label{ssec:lpt}

Lagrangian perturbation theory (LPT) models structure formation by following the trajectories of infinitesimal fluid elements from their original Lagrangian positions $\bq$ through their displacements $\Psi(\bq,t)$ , such that the observed position of each fluid element is $\bx = \bq + \Psi$ \cite{Matsubara08b,Carlson13,Porto14,Vlah15b,Vlah15}. To further transform into redshift space we insert an additional displacement equal to $\bu$, which is proportional to the time derivative of $\Psi$, thereby defining a redshift-space displacement $\Psi_s = \Psi + \bu$ \cite{Matsubara08,Carlson13}. Expanding order-by-order in the initial conditions, this is equivalent in the Einstein-de Sitter (EdS) approximation to boosting the $n^{\rm th}$ order displacement \edit{along the line-of-sight unit vector $\hat{n}$} by
\begin{equation}
    \Psi^{(n)}_{s,i} = R^{(n)}_{ij} \Psi^{(n)}_j =  \Psi^{(n)}_{i} + n f(a) \hn_i \hn_j\ \Psi^{(n)}_j
    \label{eqn:psi_s}
\end{equation}
where $R_{ij}\equiv\delta^D_{ij}+n f(a) \hn_i \hn_j$. Here $f$ is the linear growth rate $f = d\ln D(a)/ d\ln(a)$ and $\Psi^{(n)}$ denotes the perturbative displacement at that order, such that the second line-of-sight component in Equation~\ref{eqn:psi_s} is equal to the n$^{\rm th}$ order velocity $\bu^{(n)}(\bq)$. Beyond the EdS approximation the n$^{\rm th}$ order displacements and velocities are related by different factors than $n f(a)$.

Due to the equivalence principle, biased tracers like galaxies (``g'') have displacements $\Psi_g$ that follow matter on large scales, deviating on small scales only due to the small-scale physics of galaxy formation. On the other hand, the densities of biased tracers do not directly follow that of matter --- in LPT this is modeled via a bias expansion in Lagrangian coordinates \cite{Matsubara08,McDonald09,Vlah16,Desjacques18,Chen20}
\begin{equation}
    F_g(\bq) = 1 + b_\delta \delta_0(\bq) + \frac12 b_{\delta^2} \left( \delta_0(\bq)^2 - \avg{\delta_0^2} \right) + b_{s^2} \left( s_0^2(\bq) - \avg{s_0^2} \right) + b_{O_3} \mathcal{O}_3(\bq)
    \label{eqn:bias_expansion}
\end{equation}
which acts as an additional weight at Lagrangian positions $\bq$ dependent on their linear initial conditions, e.g. through the overdensity $\delta_0$ and tidal field $s_{0,ij}$. At the 1-loop order to which we will work there is one non-degenerate bias contribution at cubic order \cite{McDonald09} --- in this work we follow the convention of ref.~\cite{Chen20} and choose $O_3 = s_{ij}t_{ij}$ where $t_{ij}$ is proportional to the shear due to the second-order Lagrangian displacement and defined in ref.~\cite{McDonald09}. Advecting the bias functional to the observed redshift-space position of galaxies yields the redshift-space power spectrum \cite{Carlson13}
\begin{equation}
    P_{s}(\bk) = \int d^3 \bq\ e^{i\bk\cdot\bq} \avg{e^{i\bk\cdot\Delta_s}\ F_{g}(\bq_1)F_{g}(\bq_2) }_{\bq=\bq_2-\bq_1}.
    \label{eqn:lptpk}
\end{equation}
In the equation above, the redshift-space pairwise displacement $\Delta_s = \Psi_s(\bq_2) - \Psi_s(\bq_1)$ contains both the pairwise displacement due to particle trajectories in real space $\Delta = \Psi(\bq_2) - \Psi(\bq_1)$ and, as in Equation~\ref{equation:power_dpectrum}, due to pairwise velocities $\Delta \bu = \bu(\bx_2) - \bu(\bx_1)$. We have suppressed the counting parameter $\lambda$ in the above, though it is straightforward to include by e.g. taking $f \rightarrow \lambda f$ in Equation~\ref{eqn:psi_s}.

The expectation value in Equation~\ref{eqn:lptpk} can be worked out by expanding the exponent and Wick-contracting the initial conditions order-by-order --- indeed, this yields an identical result at any order to Eulerian perturbation theory (EPT), which we discuss next. However, it is desirable, as we will discuss below in Section~\ref{ssec:ir_resum}, to carry the contributions from linear modes in the exponent to all orders. In particular, for the linear (Zeldovich) displacement $\Psi^{(1)} = i \bk \delta_0 / k^2 $ we can use the cumulant theorem to write
\begin{equation}
    \avg{e^{i\bk\cdot\Delta^{(1)}_s}} = e^{-\frac12 k_i k_j A_{ij}^{\rm lin,s}(\bq)}, \quad A^{\rm lin, s}_{ij}(\bq) = \avg{\Delta_{s,i}^{(1)} \Delta_{s,j}^{(1)}} = R^{(1)}_{ia} R^{(1)}_{jb} \avg{\Delta_{a}^{(1)} \Delta_{b}^{(1)}}.
\end{equation}
This allows us to resum contributions that contain disconnected products with the exponential above and cast Equation~\ref{eqn:lptpk} into the form \cite{Vlah19,Chen19,Chen21}
\begin{equation}
    P_{s}(\bk) = \sum_{\substack{O_a, O_b \in \\ \{1,\delta, \delta^2, s^2, O_3\}}} b_{O_a} b_{O_b} \int d^3\bq\ e^{i\bk\cdot\bq - \frac12 k_i k_j A^{\rm lin,s}_{ij}(\bq)} f_{ab}(\bk,\bq,\hn)
    \label{eqn:clpt}
\end{equation}
where the $f_{ab}$ are Wick-contractions of the initial conditions not solely due to linear displacements $\Delta_s^{(1)}$ up to some order in perturbation theory and $b_1 = 1$. The three-dimensional integrals in Equation~\ref{eqn:clpt} have a nontrivial angular structure depending on $\hn, \bk, \bq$, such that $f_{ab}$ is a function of both their lengths and dot products $\hk\cdot\hq$, $\hn\cdot\hq$ and $\hn\cdot\hk$. We review methods to efficiently evaluate integrals of this form numerically developed in ref.~\cite{Chen21} (hereafter \textbf{C21}) in Appendix~\ref{app:lpt_integrals}. Throughout this work we will utilize the public code \texttt{velocileptors}\footnote{\url{https://github.com/sfschen/velocileptors}} released as part of that work to compute predictions for LPT power spectra.

Equation~\ref{eqn:clpt} is often called Convolutional Lagrangian Peturbation Theory (CLPT) due to the presence of a Gaussian exponent resembling a damping kernel \cite{Carlson13}. CLPT computes LPT predictions to a given order while resumming contributions from linear displacements and velocities to arbitrary order. If only long wavelength modes should be resummed, the exponent $A_{ij}^{\rm lin, s}$ can additionally be split into contributions from long and short modes $A_{ij}^{\rm lin, s} = A_{ij}^{<, s} + A_{ij}^{>, s}$, with the latter also expanded in a Taylor series up to a given perturbative order \cite{Chen20}
\begin{equation}
    P_{s}(\bk) = \sum_{a,b} b_{O_a} b_{O_b} \int d^3\bq\ e^{i\bk\cdot\bq - \frac12 k_i k_j A^{\rm <,s}_{ij}(\bq)} g_{ab}(\bk,\bq,\hn), \quad g_{ab} = e^{-\frac12 k_i k_j A_{ij}^{>,s}(\bq)} \ f_{ab}(\bk,\bq,\hn).
\end{equation}
In this case the resummed prediction at any given order is computed by keeping $g_{ab}$ to that order and adding contributions from $A_{ij}^{<}$ to all orders through the exponential. A suitable choice of splitting can tame the effects of short-wavelength linear displacements on the broadband while still capturing the smoothing of the BAO by long-wavelength displacements, as we describe in Section~\ref{ssec:ir_resum}. One well-motivated long-short split is $k_{\rm IR} = k/2$ (see e.g. the discussion in ref.~\cite{Chen24}). However, we note that the difference between theories with different long-short splits exists at 2-loop order and should be taken as part of the inherent theoretical uncertainty in the 1-loop prediction.

\subsection{Eulerian Perturbation Theory}
\label{ssec:ept}

Eulerian perturbation theory (EPT) is another common approach to analytically model structure formation (see e.g. refs.~\cite{classpt,pybird,Chen20} for well-documented public numeric implementations). Instead of following fluid-element trajectories, EPT evolves fluid densities and velocities at fixed points $\bx$ using the continuity and Euler equations. Since both EPT and LPT model the same underlying gravitational fluid, their predictions are necessarily equal order-by-order; the differences therefore boil down to the choice of bias scheme and IR resummation. The bias expansion in Eulerian coordinates can be written at 1-loop order in the power spectrum as \cite{McDonald09,Desjacques18}
\begin{equation}
    \delta_g(\bx) = c_\delta \delta(\bx) + \frac12 c_{\delta^2} (\delta^2(\bx) - \avg{\delta^2}) + c_{s^2} (s^2(\bx) - \avg{s^2}) + c_{O_3} O_3^E(\bx).
    \label{eqn:bias_parameters_ept}
\end{equation}
where $\delta(\bx)$ denotes the nonlinear matter field at the observed time, $s_{ij}(\bx)$ is the nonlinear tidal field and we again adopt $O_3^E(\bx) = s_{ij}(\bx) t_{ij}(\bx).$ The Eulerian bias parameters can be uniquely mapped onto their Lagrangian counterparts $c_{O_i} = c_{O_i}[\{b_{O_j}\}]$; the rotations specific to the conventions we use in this work are given in \textbf{C21}.

A particularly straightforward way to compute the redshift-space power spectrum in EPT is to compute the real-space pairwise velocity spectra and combine them using Equation~\ref{eqn:mome}. Calculating pairwise velocity spectra in EPT involves first computing individual real-space velocity spectra $P_{LL'}$ using nonlinear predictions for the n$^{\rm th}$ order velocity $\bu^{(n)}(\bx)$:
\begin{equation}
    P_{LL'}(\bk) = \sum_{\substack{O_a, O_b \in \\ \{1,\delta, \delta^2, s^2, O_3\}}} c_{O_a} c_{O_b} \langle (O_a \bu^L)(\bk) | (O_b  \bu^{L'})(\bk') \rangle'
\end{equation}
where again $c_1 = 1$. At 1-loop order the expansion in Equation~\ref{eqn:mome} terminates at the fourth pairwise velocity moment. These calculations were performed in detail in \textbf{C21} and implemented in \texttt{velocileptors} --- we will use these predictions extensively throughout the rest of this work and refer the interested reader to \textbf{C21} for further details.

\subsection{Effective Theory Contributions}
\label{ssec:eft_terms}

Perturbative treatments of large-scale structure are effective theories capturing the response of galaxy clustering to long-wavelength perturbations. In these theories, the effects of small-scale physics and short-wavelength (UV) modes are not computed from first principles but rather marginalized over, resulting in a finite set of free parameters at any given order whose contributions are constrained by fundamental symmetries (e.g. the equivalence principle and Galilean and rotational invariance).

In addition to the dimensionless bias parameters and gravitational displacements described in the previous subsections, it is necessary to include dimensionful effective-theory corrections whose sizes are determined by the physical scale of relevant small-scale nonlinearities, e.g. the nonlinear scale $k_{\rm nl}$, average size of intrahalo velocities $\sigma_v$ and Lagrangian halo radius $R_h$ \cite{Senatore12,Perko16,Desjacques18,Fujita20,Chen20}. 
Broadly speaking, we can categorize these corrections into: (1) \textit{counterterms}, which capture deterministic effects such as  the effective fluid stress tensor due to short-wavelength modes onto large scales,  finite nonlocalities in galaxy bias and nonlinearities in the real-to-redshift space mapping (e.g. Fingers of God) and; (2) \textit{stochastic} contributions,  which capture the additional scatter in observables due to uncorrelated short-distance physics such as shot noise in galaxies. We now briefly summarize these contributions to the redshift-space power spectrum, emphasizing terms due to different powers of the peculiar velocity ($\lambda$); we refer the interested reader to the works cited above for further details as to their derivation, and provide a rough sketch of the Lagrangian derivation in Appendix~\ref{app:counterterms}.

The counterterms  $\beta^n_m$ in the redshift-space power spectrum at 1-loop order contribute as \cite{Perko16,Chen20,Ebina24}
\begin{align}
    P_s(\bk;\lambda) \supset 2\ \big(1 + b_\delta &+ \lambda f\mu^2 \big) \nonumber \\
    \Big( \beta_0^0 &+ (\lambda f \beta_2^1 + \lambda^2 f^2 \beta_2^2) \mu^2 + (\lambda^2 f^2 \beta_4^2 + \lambda^3 f^3 \beta_4^3) \mu^4 \Big)\ k^2 P_{\rm L}(k),
    \label{eqn:counterterms}
\end{align}
where $P_{\rm L}(k)$ is the linear \textit{matter} power spectrum at a given redshift, to be distinguished from e.g. $P_{\rm lin}(\bk)$ which we use to denote the linear galaxy power spectrum.\footnote{In LPT, in order to keep long-wavelength modes exponentiated in the counterterms we substitute $P_{\rm L}$ for the leading-order $b_\delta^2$ contribution, with $f_{\delta \delta} = \xi_{\rm L}$ equal to the linear matter correlation function \cite{Maus25a}.} We have included a factor of $f^n$ in front of each $\beta^n_m$---this normalization is arbitrary but keeps the counting in $\lambda$ and $f$ identical. From the above we see there are only $5$ independent counterterms in the generating function, and indeed only $3$ in the case of the redshift-space power spectrum ($\lambda=1$). Each of these terms captures the UV dependence of many correlators of bias operators, displacements and velocities, whose effect on the pairwise velocities and therefore at each order in $\lambda$ are indistinguishable in form. As a consequence, all pairwise velocity spectra at 1-loop are described by the same set of $5$ independent counterterms, including in both real and redshift space. The sizes of these terms are expected to be at the same order as the characteristic length squared of small-scale nonlinearities, e.g. the real-space contribution ($\beta_0^0$) is of order the nonlinear scale squared $k_{\rm nl}^{-2}$ while the redshift-space ones proportional to powers of $\lambda$ can be larger at order $\sigma_v^2$. This reflects the expected effect size of the underlying UV physics, e.g. that Fingers of God due to nonlinear velocities are significant at larger scales than other dynamical nonlinearities in the redshift-space power spectrum.

We can similarly enumerate stochastic contributions to the power spectrum. Including terms up to quadratic order in the wavenumber at 1-loop we have \cite{Perko16,Chen20}
\begin{equation}
    P_s(\bk; \lambda) \supset  s_0^0 + s_0^2 k^2 + \big(\lambda f s_2^1 + \lambda^2 f^2 s_2^2 \big) (k\mu)^2 .
    \label{eqn:stoch}
\end{equation}
Here the isotropic terms have sizes governed by the interhalo separation $R_h$ ($\bar{n} = 1/R_h^3$); $s_0^0 \sim R_h^3$ and $s_0^2 \sim R_h^5$ respectively describe corrections to galaxy stochasticity from Poissonian shot noise and its scale dependence. The anisotropic terms describe additional stochasticity due to small-scale velocities and have sizes of roughly $s^1_2, s^2_2 \sim R_h^4 \sigma_v, R_h^3 \sigma_v^2$.
These contributions can be derived by examining the stochastic contributions of each (product) operator entering the redshift-space power spectrum, and again we see that there are only a limited number of contributions that describe redshift-space clustering and pairwise velocities up to 1-loop order. Indeed, we will see that the effect of both counterterms and stochastic contributions are overdetermined relative to the number of clustering multipoles up to second order in pairwise velocity statistics.

In addition to the above, the small-scale velocity dispersions of galaxies produce nonlinearities along the line of sight that can be anomalously large compared to other effects at the same naive order. This is because the typical velocity dispersion $\sigma_v$ can be much larger than the nonlinear scale $k_{\rm nl}^{-1}$ when converted into units of distance. In order to take these effects into account we can introduce the beyond 1-loop terms \cite{Ivanov20,Chen20}
\begin{equation}
    P_s(\bk; \lambda) \supset \lambda^4 f^4 \beta_{\rm FOG}\ (k \mu)^4 (b + \lambda f \mu^2)^2 P_{\rm L}(k) + \lambda^4 f^4 s_4^4 (k\mu)^4
    \label{eqn:fog_parameters}
\end{equation}
\edit{where $b = 1 + b_\delta = c_\delta$ is the linear bias}. Note that terms with this scale dependence can also enter at lower order in the velocities ($\lambda$). However, those contributions are suppressed since their coefficients will trade factors of $\sigma_v$ for $k_{\rm nl}^{-1}$, leading to a smaller effect on large scales. Supposing that the leading contributions beyond 1-loop order come from stochastic velocities $\epsilon_i$ and ignoring their correlation with any other deterministic or stochastic contributions gives an expansion in powers of $\sigma_v k \mu$ multiplying the rest of the power spectrum as in traditional models of Fingers of God.

\subsection{IR Resummation of the BAO in EPT}
\label{ssec:ir_resum}

We conclude our review of the power spectrum with a quick summary of the infrared (IR) resummation of the BAO, particularly in Eulerian perturbation theory. Briefly, the presence of an additional scale in the power spectrum, i.e. a well-localized peak in the linear correlation function due to baryon acoustic oscillations at $r_d \approx 150 \text{Mpc}$, means that our perturbative predictions are sensitive to an additional parameter $k \Sigma_d$ where \cite{Baldauf15,Blas16,Vlah16b,Chen24}
\begin{equation}
    \Sigma^2_d = \frac23 \int \frac{dk}{2\pi^2} (1 - j_0(k r_d) + 2 j_2(k r_d)) P_{\rm L}(k)
\end{equation}
is the variance of pairwise linear displacements $\Delta^{(1)}$ smearing the BAO, evaluated at Lagrangian radii $q = r_d$. At redshift $z = 0$, $(k \Sigma_d) > 1$ for $k > 0.1 h^{-1}\text{Mpc}$, such that the effects of this additional parameter are not perturbative even on these scales. In order to properly model the effect of the long-wavelength modes contributing to $\Sigma_d$ on the BAO feature,  it is necessary to resum their contributions, i.e. include their effects beyond a given order in perturbation theory. Indeed, the bulk of the contributing displacements live in the linear regime, making it possible to exactly capture their nonlinear effects to all orders. Let us briefly review the derivation for these non-perturbative effects following the approach in ref. \cite{Vlah16b}.

From Equation~\ref{eqn:lptpk}, the power spectrum in LPT can be Taylor-expanded in terms of pairwise displacements as
\begin{equation}
    P_s(\bk) = \sum_{n=0}^\infty \frac{i^n k_{i_1} \ldots k_{i_n}}{n!}\avg{ (1 + \delta_1) (1 + \delta_2) \Delta_{s, i_1} \ldots \Delta_{s, i_n}}(\bk).
\end{equation}
To perform IR resummation we want to pull out disconnected contributions from linear displacements $\Delta^{(1)}$, i.e.\footnote{Here we are dropping contributions from higher-order $\Delta^{(n)}$, which can be absorbed into $\delta_{1,2}$ for the sake of this argument.} 
\begin{equation}
    \avg{ (1 + \delta_1) (1 + \delta_2) \Delta_{s, i_1} \ldots \Delta_{s, i_n}} \supset \avg{ (1 + \delta_1) (1 + \delta_2)} \avg{\Delta_{s, i_1}^{(1)} \ldots \Delta_{s, i_n}^{(1)}}
    \label{eqn:pk_diagram}
\end{equation}
where the latter factor can be easily computed as the product of $n/2$ pairs using Wick's theorem and the approximation \cite{Chen24}
\begin{equation}
    k_i k_j \avg{\Delta_{s, i_1}^{(1)}  \Delta_{s, i_n}^{(1)}}_{q = r_d} \approx k^2 \Sigma^2_d(\mu), \quad \Sigma^2_d(\mu) = (1 + f(2+f) \mu^2)\ \Sigma^2_d.
\end{equation} 
The point of IR resummation is to pull out the effect of this factor on the BAO, given that the BAO in the former factor $\avg{ (1 + \delta_1) (1 + \delta_2)}$ manifests as a sharp peak at $q = |\bq_1 - \bq_2| = r_d$. In this case, we can split the linear spectrum $P_{\rm L}$ into wiggle ($P^w_{\rm L}$ ) and no-wiggle ($P^{nw}_{\rm L}$) contributions, i.e.  $P_{\rm L} = P^{w}_{\rm L} + P^{nw}_{\rm L}$, where the Fourier transform of the former only has support about the peak. Similarly,  we can split \textit{nonlinear} correlators like the power spectrum $P_s[P_{\rm L}]$ into no-wiggle and wiggle contributions
\begin{equation}
    P_{s,w} = P_s[P_{\rm L}] - P_s[P_{\rm L}^{nw}]
    \label{eqn:wiggle_contribution_def}
\end{equation}
by computing the nonlinear statistic with or without wiggles.

In particular, since the wiggle component of the former factor in Equation~\ref{eqn:pk_diagram},  i.e. $\avg{ (1 + \delta_1) (1 + \delta_2)}$,   is sharply localized in configuration space at $q = r_d$, we can substitute the smooth latter factor, $\avg{\Delta_{s, i_1}^{(1)} \ldots \Delta_{s, i_n}^{(1)}}$, for its value at $r_d$ when multiplying it, such that we have the wiggly contributions \cite{Vlah16b}
\begin{equation}
    \frac{i^n}{n!} k_{i_1} \ldots k_{i_n} \avg{\Delta_{s, i_1} \ldots \Delta_{s, i_n}}_{q=r_d} \avg{ (1 + \delta_1) (1 + \delta_2)}_w (\bk) \approx \frac{1}{(n/2)!} \llbracket P(\bk) \rrbracket_w \left(\frac{-k^2 \Sigma^2_d(\mu) }{2} \right)^{(n/2)}.
    \label{eqn:resummed_pk_diagram}
\end{equation}
Summing up contributions gives us that the wiggly components are given by \cite{Ivanov18}
\begin{align}
    P^{\rm 1loop}_w(\bk) &= e^{-\frac12 k^2 \Sigma^2(\mu)}\ \llbracket P(\bk) \rrbracket^{\rm 1loop}_w \nonumber \\
    &= e^{-\frac12 k^2 \Sigma^2(\mu)}\ \left( P^{\rm lin}_w(\bk) + P^{\rm loop}_w(\bk) + \frac12 k^2 \Sigma^2_{d}(\mu) P^{\rm lin}_w(\bk) \right).
\end{align}
In the above we have defined $\llbracket P(\bk) \rrbracket_w^{\rm 1loop}$ to be the contributions to the (wiggly) power spectrum at a 1-loop order subtracting those with disconnected contributions by long-wavelength displacements $\Sigma_d$. For example, at $n = 2$,  Equation~\ref{eqn:resummed_pk_diagram} contains $- k^2 \Sigma_d^2(\mu) P^{\rm lin}_w/2$, which is formally at 1-loop order but is captured by IR resummation as part of $e^{-\frac12 k^2 \Sigma_d^2(\mu)} P^{\rm lin}_w$. Since this contribution is resummed we have to subtract it from $\llbracket P \rrbracket_w$ to avoid double counting. If, for example, we instead want to perform IR resummation on the \textit{linear} theory prediction, we would instead have $\llbracket P \rrbracket_w^{\rm linear} = P_w^{\rm lin}(\bk)$, since there are no disconnected contributions involving $\Sigma^2_d$ at that order. Isolating contributions to the wiggly power spectrum in the absence of long-wavelength displacements will be critical to constructing an IR resummation scheme for velocity statistics, as we will show below. 

It is important to note that since we have derived these expressions from LPT, these effects are naturally included in LPT calculations while they have to be added by hand into EPT calculations. This is because the exponential factor in Equation~\ref{eqn:clpt} naturally captures this smoothing effect, which can indeed also be directly derived by taking a saddle point approximation of that factor at $q = r_d$ \cite{Vlah16b}.

\section{Pairwise Velocity Statistics in Perturbation Theory}
\label{sec:pv_pt}

We are now in a position to compute effective-theory predictions for pairwise peculiar velocity statistics in both the Lagrangian and Eulerian frameworks. Indeed, these predictions follow directly from the derivative formula (Eqn.~\ref{eqn:generating_function}), though they involve both additional subtleties and some simplifications as we will now describe.

\subsection{LPT}
\label{ssec:lpt_pv}
\subsubsection{Two Routes to Peculiar Velocities in LPT}
There are two obvious methods to compute redshift-space pairwise velocity spectra in LPT --- one can take numerical derivatives of Equation~\ref{eqn:lptpk} directly or, alternatively, work out the contributions to each pairwise velocity moment by taking analytic derivatives of the same expression. The latter method is formally equivalent to computing
\begin{equation}
    \tilde{\Xi}^{(n)}_{s}(\bk) = \int d^3 \bq\ e^{i\bk\cdot\bq} \avg{e^{i\bk\cdot\Delta_s}\ \Delta \bu^n_{\hn}\ F_{g}(\bq_1)F_{g}(\bq_2) }_{\bq=\bq_2-\bq_1}
    \label{eqn:xin_lpt}
\end{equation}
whose real-space counterpart was explored in refs.~\cite{Vlah16,Chen20}.
These two methods are formally equivalent up to any order in perturbation theory, but we need to be slightly careful in order to enforce this equivalence in practical calculations, specifically with regard to the terms we resum.

Let us visit both methods by examining the statistics of matter, i.e. an unbiased tracer with $F_g = 1$ in Equation~\ref{eqn:lptpk}. In this case at 1-loop in perturbation theory we have \cite{Carlson13,Vlah15}
\begin{equation}
    P_{s}(\bk) = \int d^3\bq\ e^{i\bk \cdot \bq - \frac12 k_i k_j A_{ij}^{\rm lin, s}(\bq)} \left(1 - \frac12 k_i k_j A_{ij}^{\rm loop, s}(\bq) - \frac{i}{6} k_i k_j k_k W_{ijk}^{\rm loop, s}(\bq) \right)
    \label{eqn:matter_pk_lpt}
\end{equation}
where
\begin{align}
    A_{ij}^{\rm loop, s} &= R^{(2)}_{ia} R^{(2)}_{jb} A^{(22)}_{ab} + R^{(1)}_{ia} R^{(3)}_{jb} A^{(13)}_{ab} + (1 \leftrightarrow 3) \nonumber \\
    W_{ijk}^{\rm loop, s} &= R^{(1)}_{ia} R^{(1)}_{jb} R^{(2)}_{kc} W^{(112)}_{abc} + (112 \leftrightarrow 121) + (112 \leftrightarrow 211)
\end{align}
and $A^{(nm)} = \avg{\Delta^{(n)}\Delta^{(m)}}$, $W^{(nmo)} = \avg{\Delta^{(n)}\Delta^{(m)}\Delta^{(o)}}$ are perturbative contributions to the second and third moments of the pairwise displacement $\Delta$. The $\lambda$ dependence in the above is encoded in the redshift-space matrices $R^{(n)}$, with $\partial_\lambda R^{(n)}_{ij} = n f \hn_i \hn_j$, so that computing higher-order pairwise velocity statistics amounts to projecting displacement statitstics like $A_{ij}, W_{ijk}$ along the line of sight. Note that in the above discussion we have not included (necessary) counterterms and stochastic terms, whose $\lambda$ derivatives must be taken explicitly following Equations~\ref{eqn:counterterms} and \ref{eqn:stoch}.

In fact, CLPT predictions for pairwise velocity statistics in real space were presented in \textbf{C21}, and their extensions to redshift space simply involve mapping $\Psi \rightarrow \Psi_s$ using the matrices $R^{(n)}$ while keeping the velocities unchanged $\dot{\Psi} \rightarrow \dot{\Psi}$.  In this scheme, all terms are expanded to a given (e.g. 1-loop) order except for the linear $A_{ij}$ in the exponent. For example, the first moment of the pairwise velocity for matter is given by
\begin{equation}
    v^{\bf C21}_{s}(\bk) = \int d^3\bq\ e^{i\bk \cdot \bq - \frac12 k_i k_j A_{ij}^{\rm lin, s}(\bq)} \left( i k_j \dot{A}^{\rm s}_{j\hn} - \frac12 k_j k_k \dot{W}^{\rm s}_{jk\hn} \right)
    \label{eqn:vk_direct_lpt}
\end{equation}
where we have defined $\dot{A}^{\rm s} = \avg{\Delta_s \dot{\Delta}}$, $\dot{W^{\rm s}} = \avg{\Delta_s \Delta_s \dot{\Delta}}$, equivalent to Equation 4.13 in \textbf{C21} except with undotted displacements boosted to redshift space. The numerical evaluation of this expression is straightforward since the line-of-sight projections leave $v_{s}$ in the same form as Equation~\ref{eqn:clpt}; we give some relevant details, including the extension to biased tracers, in Appendix~\ref{app:lpt_integrals}.

In Equation~\ref{eqn:vk_direct_lpt} we have kept all the contributions not in the resummed exponent to 1-loop order. However, this would not have been the result if we had directly taken the $\lambda$ derivative of Equation~\ref{eqn:matter_pk_lpt}, which would've additionally yielded the two-loop contributions
\begin{equation}
    v_{s}(\bk) \supset \int d^3\bq\ e^{i\bk \cdot \bq - \frac12 k_i k_j A_{ij}^{\rm lin, s}(\bq)}\  i k_j \dot{A}^{\rm lin, s}_{j\hn}(\bq) \left( - \frac12 k_i k_j A_{ij}^{\rm loop, s}(\bq) - \frac{i}{6} k_i k_j k_k W_{ijk}^{\rm loop, s}(\bq) \right).
    \label{eqn:2loop_diff}
\end{equation}
Since each $\lambda$ derivative of the exponent brings down an additional power of $A_{ij}$, applying the derivative formula to compute $v_s$ and $\sigma^2_s$ will produce differences at two loops and above when compared to the formalism in \textbf{C21}. 

\subsubsection{IR-Resummed Peculiar Velocties in LPT: General Method}

More generally, from Equation~\ref{eqn:clpt} we can write the generating function (power spectrum) in the form
\begin{equation}
    P_s(\bk;\lambda) = \int d^3\bq\ e^{i\bk\cdot\bq - \frac12 k_i k_j A^{\rm lin, s}_{ij}(\bq)} [ \tilde{P}_s ]  (\bq;\lambda)
    \label{eqn:brackets_lpt}
\end{equation}
where we have defined $[\tilde{P}_s]$ to be the Fourier-transformed contributions to the power spectrum integrand that do not contain disconnected products of the pairwise-displacement 2-point function $A_{ij}^{\rm lin}$, e.g. the terms in parentheses in Equation~\ref{eqn:matter_pk_lpt} in the case of the 1-loop matter power spectrum, or more generally $\sum_{O_a,O_b} b_{O_a} b_{O_b} f_{ab}$ in Equation~\ref{eqn:clpt}. When doing the long-short split in linear displacements $A_{ij}^{\rm lin}$ should be replaced by $A_{ij}^{<}$. Keeping these terms to a given, e.g. 1-loop, order while resumming the pairwise-displacement two-point function is achieved by using the derivative formula, e.g.
\begin{equation}
    (ik\mu) v_s(\bk) = \int d^3\bq\ e^{i\bk\cdot\bq - \frac12 k_i k_j A^{\rm lin, s}_{ij}(\bq)} \left( -\frac12 k_i k_j (\partial_\lambda A_{ij}^{\rm lin, s}(\bq))  [\tilde{P}_s ](\bq;\lambda) +  (ik\mu) [\tilde{v}_s ] (\bq;\lambda) \right)
\end{equation}
where we can recursively compute the contributions to the n$^{\rm th}$ velocity moment that do not contain disconnected $A_{ij}^{\rm lin}$'s as e.g. $\partial_\lambda [\tilde{P}_s] = (ik\mu) [\tilde{v}_s]$, $\partial^2_\lambda [\tilde{P}_s] = (ik\mu)^2 [\tilde{\sigma^2}_s]$, etc. If all the bracketed quantities in the parentheses are kept to 1-loop order, we can regard this as a resummed 1-loop prediction for $v_s$. On the other hand, in the CLPT prescription where only the exponent is kept resummed, some of the terms in parentheses are dropped, e.g. those contained in the first term corresponding to Equation~\ref{eqn:2loop_diff} in the case of matter. This is because their formal counting \textit{with} long wavelengths included is beyond 1-loop order. We will refer broadly to this latter scheme---and its counterpart in EPT---where only the exponent is resummed as the \textbf{C21} scheme.  Naively, introducing higher-order terms beyond the \textbf{C21} scheme while neglecting others may seem inconsistent with the spirit of perturbation theory, though as we will explore in more detail below these additional terms are actually \textit{required} for a consistent counting of the large paramter $k \Sigma_d$ and critical in correctly modeling baryon acoustic oscillations in velocity statistics. Note that a nearly identical derivation can be performed in the case of real-space velocity statistics, except with setting $\lambda = 0$ at the end, yielding a different resummation scheme than that in \textbf{C21} (Appendix~\ref{app:ir_resummation_real_space}).

In practice, to evaluate velocity statistics up to second order in the former method involves taking a central difference by evaluating the redshift-space power spectrum at $\lambda = 1, 1 \pm \epsilon$ and at up to two angular ($\mu$) positions in order to evaluate angular multipoles up to the quadrupole. In comparison, evaluating the same multipoles through direct Lagrangian evaluation after taking analytic derivatives up to second order in the pairwise velocities involves computing 3 spectra at an equivalent number of angular positions, making both methods roughly equivalent in terms of computational complexity.

\subsection{EPT}

\subsubsection{Peculiar Velocities in EPT via the Moment Expansion}

The calculation of pairwise velocity statistics in (un-resummed) EPT is much more straightforward. Using the moment expansion (Eqn.~\ref{eqn:mome}) we have that \cite{Okumura14}
\begin{equation}
    \tilde{\Xi}^{(m)}_s(\bk) = \sum_{n=0}^\infty \frac{(ik\mu)^n}{n!} \tilde{\Xi}^{(n+m)}(\bk).
\end{equation}
At 1-loop order the real-space pairwise velocity moments are nonzero up to $\tilde{\Xi}^{(4)}$, so we can write e.g.\footnote{We connect this expression for the first pairwise velocity moment to a more common one for the galaxy density-momentum cross correlation based on non-pairwise peculiar velocity spectra \cite{Okumura14} in Appendix~\ref{app:okumura}, correcting a small typo in previous works and showing that they are equivalent.}
\begin{equation}
    v_{s}^{\rm 1loop}(\bk) = v^{\rm 1loop}(\bk) + (ik\mu) \sigma^{2,\rm 1loop}(\bk) + \frac{(ik\mu)^2}{2} \tilde{\Xi}^{(3), \rm 1loop}(\bk) + \frac{(ik\mu)^3}{6}\tilde{\Xi}^{(4), \rm 1loop}(\bk).
    \label{eqn:mome_vk}
\end{equation}
The EPT predictions for these real-space pairwise velocity statistics were derived in \textbf{C21} and implemented in \texttt{velocileptors}, so it is straightforward to apply them to compute their redshift-space counterparts. However, this unresummed treatment does not take into account the non-perturbatvely large effects of long displacements on the BAO, which we turn to now.

\subsubsection{Recursive IR Resummation Scheme for Pairwise Velocities in EPT}
\label{ssec:ir_resum_pv}

In this section we will derive an IR resummation scheme appropriate for pairwise velocities using similar logic to Section~\ref{ssec:ir_resum}. While we will focus on the case of redshift-space peculiar velocity statistics, the same considerations below also imply an improved resummation scheme in real space compared to earlier work, which we describe in Appendix~\ref{app:ir_resummation_real_space}.

Let us begin with the first pairwise velocity moment, which we can again write in LPT via Equation~\ref{eqn:xin_lpt} as
\begin{equation*}
    v_s(\bk) = \sum_{n=0}^\infty \frac{i^n k_{i_1} \ldots k_{i_n}}{n!}\avg{ (1 + \delta_1) (1 + \delta_2) \dot{\Delta}_\parallel \Delta_{s, i_1} \ldots \Delta_{s, i_n}}(\bk).
\end{equation*}
We are interested in contributions where linear displacements $\Delta^{(1)}$ are contracted only against each other. Unlike in the power spectrum, here we have two types:
\begin{align}
    & \avg{ (1 + \delta_1) (1 + \delta_2) \dot{\Delta}_\parallel \Delta_{s, i_1} \ldots \Delta_{s, i_n}} \nonumber \\
     & \;\supset \avg{ (1+\delta_1)(1+\delta_2)} \avg{ \dot{\Delta}_\parallel^{(1)} \Delta_{s, i_1}^{(1)} \ldots \Delta_{s, i_n}^{(1)}}, \; \avg{ (1+\delta_1)(1+\delta_2) \dot{\Delta}_\parallel^{(1)} } \avg{ \Delta_{s, i_1}^{(1)} \ldots \Delta_{s, i_n}^{(1)}}.
     \label{eqn:contractions}
\end{align}
The second kind of term gives rise exactly to the usual Gaussian damping of wiggles and is the equivalent of Equation~\ref{eqn:pk_diagram} for the power spectrum.
However, the first one is different, since in this case the velocity we are correlating interacts with the underlying displacements. Indeed we can see that
\begin{equation}
   \langle \dot{\Delta}_\parallel \underbrace{ \Delta_s \ldots \Delta_{s}}_{n \text{ } \Delta\text{'s}}\rangle = n \times \left( \frac{(n-1)!}{2^{\frac{n-1}{2}} (\frac{n-1}{2})!} \right) \avg{ \dot{\Delta}_\parallel \Delta_{s}} \underbrace{\avg{\Delta_s \ldots \Delta_{s}}}_{(n-1) \text{ } \Delta\text{'s}} \sim n! \times \left( \frac{f k\mu (1 + f) \Sigma^{2n}_d}{2^{\frac{n-1}{2}} (\frac{n-1}{2})!} \right) \nonumber 
\end{equation}
such that, summing over all $n$, we get a contribution equal to 
\begin{equation}
    i f k\mu (1+f) \Sigma_d^2\ e^{-\frac12 k^2 \Sigma^2_d(\mu)}  \llbracket P(\bk) \rrbracket_w
    \label{eqn:new_contribution_vk}
\end{equation}
where we again note that the expectation value of the densities is taken without disconnected contributions from linear displacements. 

From the above derivation we have that the full wiggle contribution\edit{, defined as in the power spectrum in Equation~\ref{eqn:wiggle_contribution_def},} is
\begin{align}
    &v_{s,w}(\bk) = e^{-\frac12 k^2 \Sigma^2_d(\mu)} \llbracket v_s(\bk) \rrbracket_w + i f k\mu (1+f) \Sigma_d^2\ e^{-\frac12 k^2 \Sigma^2_d(\mu)} \llbracket P_s(\bk) \rrbracket_w.
    \label{eqn:resummed_vw}
\end{align}
As in the power spectrum the connected components must be constructed at each order in pairwise velocities, for example at 1-loop we need to remove two contributions in Equation~\ref{eqn:contractions}:
\begin{equation}
    \llbracket v_s(\bk) \rrbracket_w^{\rm 1loop} = v^w_{s,\rm lin}(\bk) + v^w_{s,\rm loop}(\bk) + \frac12 k^2 \Sigma^2(\mu) v^w_{s,\rm lin}(\bk) - i f(1+f) k\mu \Sigma^2_d P^w_{s,\rm lin}(\bk).
\end{equation}
while at linear order $\llbracket v \rrbracket_w = v_w^{\rm lin}$. It is easy to see that this resummation scheme follows directly from the derivative formula using $d\Sigma^2(\mu)/d\lambda = 2f(1+f)\mu^2 \Sigma^2_d$ and $d\llbracket P \rrbracket/d\lambda = ik\mu \llbracket v \rrbracket$. This is a consequence of the resummed nonlinear power spectrum being the generator of the higher moments. There is a close correspondence to the LPT quantities $[\ldots]$ defined in the discussion around Equation~\ref{eqn:brackets_lpt} with the connected EPT contributions $\llbracket \ldots \rrbracket$. Indeed, in the case of 1-loop matter this term corresponds precisely to the contribution of Equation~\ref{eqn:2loop_diff} to the BAO, where $i k_j \dot{A}_{ji}$ evaluates to $i f k \mu (1 + f) \Sigma^2_d$ and the 1-loop term in parentheses to $\llbracket P \rrbracket_w$, suggesting that this resummation we are describing corresponds to the derivative scheme in LPT. Previous models of the velocity spectra (e.g.~\textbf{C21}) have typically resummed only the diagrams leading to an exponential damping, e.g. at 1-loop
\begin{equation}
    v_{s,w}^{\rm C21}(\bk) = e^{-\frac12 k^2 \Sigma^2_d(\mu)} \{ v_s(\bk) \}_w, \quad \{ v_s(\bk) \}_w = v^w_{s,\rm lin}(\bk) + v^w_{s,\rm loop}(\bk) + \frac12 k^2 \Sigma^2(\mu) v^w_{s,\rm lin}(\bk)
\end{equation}
neglecting the also-enhanced diagrams multiplying $\llbracket P \rrbracket_w$.

We can take a similar approach to the second moment. Taking the second derivative we get at 1-loop
\begin{align}
    \sigma^{2, \rm 1loop}_{s,w}(\bk) 
    &= e^{-\frac12 k^2 \Sigma^2_d(\mu)} \Big( \llbracket \sigma^2_s(\bk) \rrbracket^{\rm 1loop}_w + 2f(1+f) \Sigma^2_d (ik\mu) \llbracket v_s(\bk) \rrbracket^{\rm 1loop}_w \nonumber \\
    &\quad \quad \quad \quad \quad \quad \quad \quad \quad \quad \quad \quad \quad + \big( (ik\mu)^2 f^2(1+f)^2  \Sigma^4_d + f^2 \Sigma^2_d \big) \llbracket P_s \rrbracket^{\rm 1loop}_w \Big) \nonumber
\end{align}
where, noting that some of the disconnected contributions above are naively at 2-loop order or higher:
\begin{align}
    \llbracket \sigma^2_s(\bk) \rrbracket^{\rm 1loop}_w = \big( 1 &+ \frac12 k^2 \Sigma^2_d(\mu) \big) \sigma_{s,\rm lin}^w(\bk) + \sigma_{s,\rm loop}^w(\bk) \nonumber \\
    &- 2 (ik\mu) f (1 + f) \Sigma^2_d v_{s,\rm lin}^w(\bk) - f^2 \Sigma^2_d P_{s,\rm lin}^w(\bk).
\end{align}
It is easy to see that this corresponds to separating out the disconnected pieces when evaluating Wick contractions of only linear displacements. Importantly, we note that the expectation values $\llbracket \ldots \rrbracket$ can be recursively constructed, e.g. $d^2\llbracket P \rrbracket / d\lambda^2 = (i k \mu)^2 \llbracket \sigma^2 \rrbracket$, in the sense that at each order they require only their counterparts at lower orders, such that the resummation scheme at higher orders can be straightforwardly constructed. This is because the n$^{\rm th}$ moment with the long displacement contributions removed is simply the n$^{\rm th}$ derivative of the power spectrum with long displacement contributions removed.

\begin{figure}
    \centering
    \includegraphics[width=\textwidth]{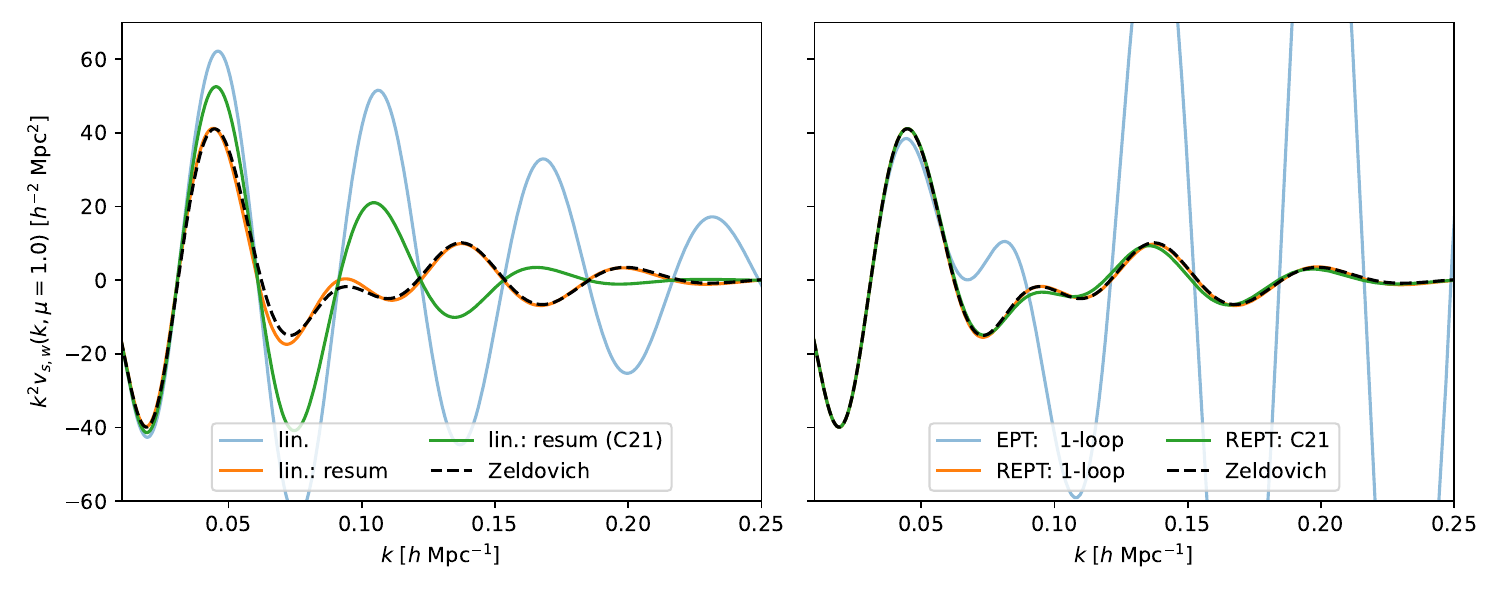}
    \caption{BAO in the $z = 0$ pairwise velocity spectrum in the Zeldovich approximation for a tracer with linear Lagrangian bias $b_\delta = 0.5$, evaluated at an angle $\mu = 1$ parallel to the line of sight. (Left) Comparison of the exact Zeldovich predction (black dashed) to the prediction of various resummation schemes. Both linear theory and the \textbf{C21} resummation scheme deviate from Zeldovich at fairly large scales while the new resummation scheme is in good agreement. (Right) The same comparison with PT predictions at 1-loop order. The \textbf{C21} resummation is notably improved though still in worse agreement than the new scheme.}
    \label{fig:vk_ir_resum}
\end{figure}

\section{Peculiar Velocities in the Zeldovich Universe}
\label{sec:pv_zel}

In order to better understand and test the various resummation schemes above let us consider a toy ``Zeldovich universe'' where structure formation is completely described by the Zeldovich approximation, i.e. $\Psi = \Psi^{(1)}$. This toy model exactly captures the effects of linear, long-wavelength modes on the BAO and, in addition, since the displacements are Gaussian, we can exactly compute pairwise statistics up to any order using the numerical LPT methods described above. In particular, since we can exactly compute the redshift-space power spectrum \cite{DeRose23}, these statistics are simply given by the derivative formula. Indeed, this property was exploited by ref.~\cite{Sugiyama16}, who used it to predict the smearing of the BAO peak in peculiar velocity statistics in configuration space, finding that the Zeldovich approximation is able to explain the reduced amplitude of the BAO feature seen in N-body simulations. It is also straightfoward to compute these statistics in an ``Eulerian'' fashion by perturbatively expanding order-by-order in the initial conditions --- one practical method to do so is to set the resummation scale in the IR-UV split in the exponentiated $A_{ij}$ to $k_{\rm IR} \rightarrow 0$, such that all linear displacements are expanded rather than resummed. Similarly, we can also evaluate LPT predictions in the \textbf{C21} scheme using the techniques described in Section~\ref{ssec:lpt_pv} and drop terms that are included in the recursive scheme. \edit{Throughout this section we will evaluate predictions in the cosmology of the N-body simulations described in Section~\ref{sec:nbody} at redshift $z=0$ for a tracer with linear Lagrangian bias $b_\delta = 0.5$.}

Let us begin by comparing velocity statistics in the Zeldovich universe to those predicted by different resummation schemes in EPT, where it is especially easy to interpret the differences. Figure~\ref{fig:vk_ir_resum} shows the BAO predictions of these schemes compared to the exact Zeldovich calculation (black dashed) for the first pairwise velocity moment $v_{s}$.   To show the strongest possible difference between different schemes we have chosen a line-of-sight angle $\mu = 1$ where redshift-space distortions are most magnified. The left panel shows the effect of different resummation schemes on linear theory, where $\llbracket\ldots\rrbracket$ are simply given by their linear predictions. The exponential-only resummation (green), i.e. that of \textbf{C21}, performs only marginally better than unresummed linear theory, while the resummation scheme proposed above in this work (orange) is able to match both the phase and amplitude of the BAO oscillations quite well, even when they depart significantly from linear-theory expectations. Similarly, the right panel shows the effect of different resummation schemes at 1-loop order. In this case both the \textbf{C21} scheme and the scheme proposed in this work are able to qualitatively describe the Zeldovich curve rather well, though the new scheme still provides a visibly better match, including capturing the phase shift visible at large $k$.

The oscillatory structure of the Zeldovich-universe BAO in Figure~\ref{fig:vk_ir_resum} is at first glance quite strange---it very quickly deviates from linear theory even on fairly large scales, then appears to halve its wavelength around $k = 0.1 h$ Mpc$^{-1}$ before becoming completely out-of-phase at higher $k$. We can understand this by computing the effective window modulating the velocity spectrum BAO $v_{s,w}(\bk) = W_v(k,\mu) P_{\rm L}(k)$. In linear theory this is simply given by the Kaiser prediction with linear bias ($b = c_\delta = 1 + b_\delta$)
\begin{equation}
    W_{v,\rm lin}(k,\mu) = - \frac{2 f \mu (b + f\mu^2)}{k}.
\end{equation}
Evidently this window is monotonic in $k$ and cannot accomodate the behavior described above. Indeed, neither can the window in the \textbf{C21} scheme, 
\begin{equation}
W_{v,\rm lin}^{\rm C21}(k,\mu) = W_{v,\rm lin}(k,\mu) e^{-\frac12 k^2 \Sigma^2(\mu)}.
\end{equation}
However, plugging in the linear-theory predictions for $[[v_w]], [[P_w]]$ into Equation~\ref{eqn:resummed_vw} instead yields the effective window
\begin{equation}
    W_{v,\rm lin}^{\rm res}(k,\mu) = - \left( \frac{2 f \mu (b + f\mu^2)}{k} \right) \left( 1 - \frac{1}{2f} (k \Sigma_d)^2 (b + f\mu^2) \right) e^{-\frac12 k^2 \Sigma^2(\mu)}.
    \label{eqn:vw_window}
\end{equation}
This window has the important property that it crosses zero and changes sign at $k \approx 0.1 h$ Mpc$^{-1}$ at $z=0$, producing an additional node in the velocity spectrum BAO and flipping the phase at high $k$. The fact that $k \Sigma_d \gtrsim 1$ even on fairly large scales makes the second term crucial to match the shape of the BAO even in the linear regime, and we note in addition that this effect is particularly pronounced for highly biased tracers with $b \gg f$.

\begin{figure}[t]
    \centering
    \includegraphics[width=\textwidth]{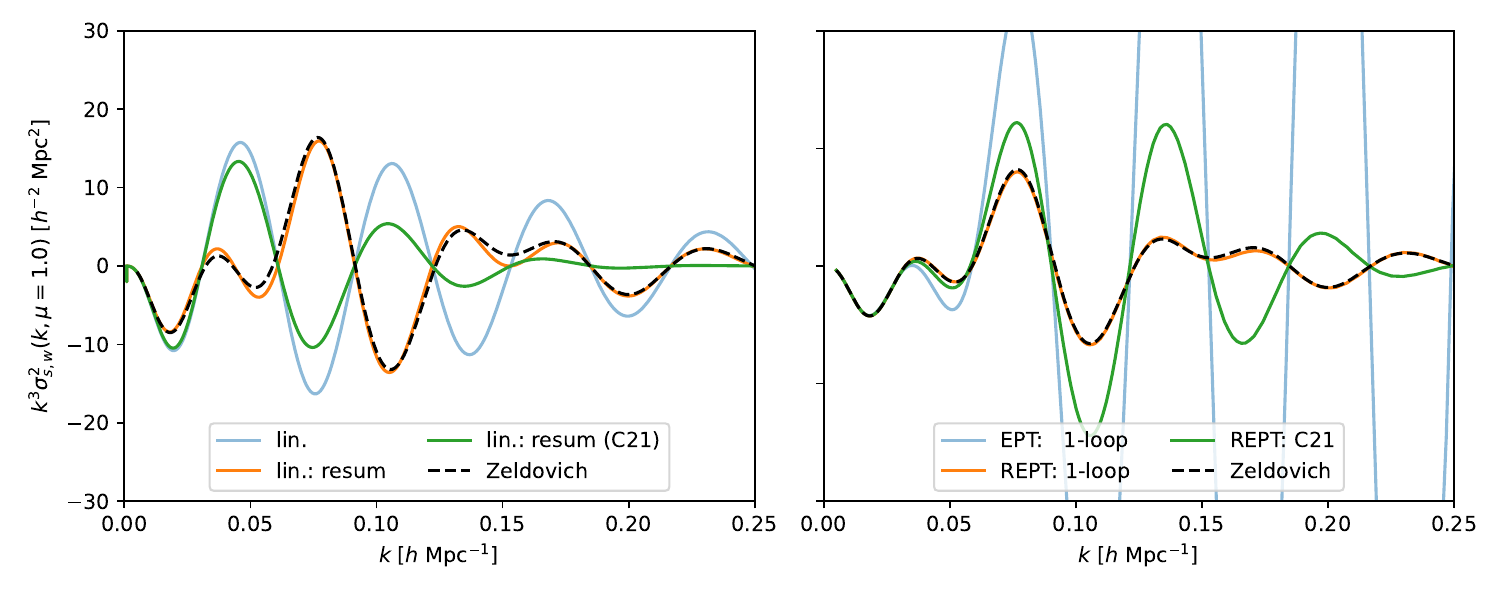}
    \caption{Same as Figure~\ref{fig:vk_ir_resum} but for the pairwise velocity dispersion spectrum $\edit{\sigma^2_s(k)}$. Notably, the 1-loop prediction in the \textbf{C21} scheme now sigificantly overpredicts the BAO amplitude even at $k < 0.1\ h$ Mpc$^{-1}$ while the new scheme is in excellent agreement with the full Zeldovich calculation at both linear and 1-loop order.}
    \label{fig:sk_ir_resum}
\end{figure}

We show similar predictions for the second pairwise velocity moment in Figure~\ref{fig:sk_ir_resum}. The differences between the IR-resummation schemes persists at this order: the left panel shows that in this case the \textbf{C21} scheme fails to capture even the sign of the BAO feature at scales as large as $k = 0.05 h$ Mpc$^{-1}$, while the new scheme in this work faithfully describes BAO oscillations out to even fairly small scales. Similarly to the velocity spectrum we can compute the resummed linear theory window
\begin{align}
    W_{\sigma^2,\rm lin}^{\rm res}(k,\mu) = -\frac{2 f^2 \mu^2}{k^2} \Big(1 - \big( 2 (1 + f ) (b &+ f\mu^2)  \nonumber \\
    + \frac{1}{2\mu^2} (b + f\mu^2)^2 \big) (k \Sigma_d)^2 
    &+ \frac12 (1 + f)^2 (b + f\mu^2)^2 (k\Sigma_d)^4 \Big)\ e^{-\frac12 k^2 \Sigma^2(\mu)}
\end{align}
where we now have a quartic function in $k \Sigma_d$ with two poles. These additional nodes and sign flips are clearly visible in Figure~\ref{fig:sk_ir_resum}.
The differences are even more pronounced at 1-loop order, where a pure exponential damping significantly overpredicts the amplitude of the BAO feature at all scales shown, while the new scheme at 1-loop order improves upon the resummed linear-theory prediction. These numerical experiments thus suggest that resumming contributions to the BAO beyond a simple exponential damping are important for pairwise velocity statistics. 

\begin{figure}
    \centering
    \includegraphics[width=\textwidth]{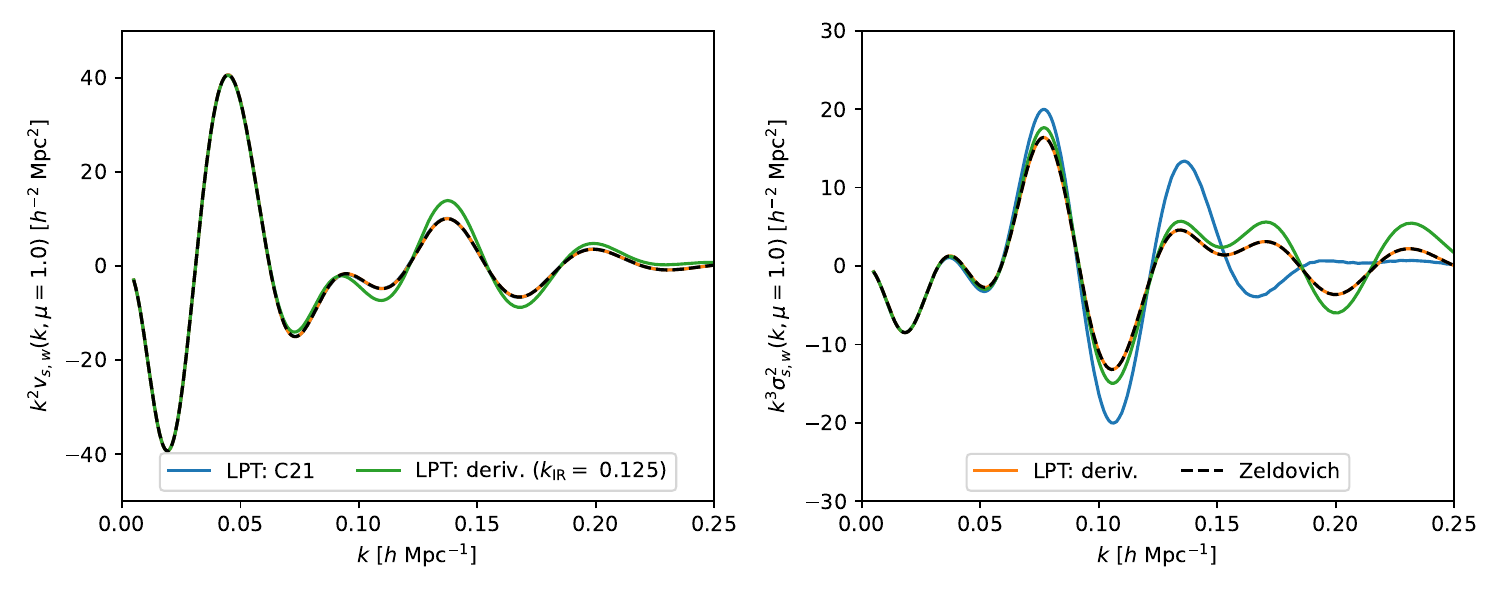}
    \caption{Predictions for the BAO contributions to $v_s$ and $\sigma^2_s$ in the Zeldovich universe using LPT. Here, the derivative formula is exact, such that the black-dashed and the orange lines are identical. On the other hand, the \textbf{C21} scheme (blue) shows significant differences in its prediction for $\sigma^2_s$, though it succeeds in predicting $v_s$ almost exactly. Predictions for resummed LPT with only modes longer than  $k_{\rm IR} = 0.125 h$ Mpc$^{-1}$ resummed in the derivative scheme is shown in green, showing good quantitative agreement on perturbative scales while differing towards smaller scales where the UV modes effects become more important. }
    \label{fig:lpt_ir_resum}
\end{figure}

Beyond EPT, the absence of these relevant terms in the ``direct'' \textbf{C21} LPT calculation to 1-loop order suggests that the recursive scheme is necessary to properly model the BAO in LPT. Indeed, we can also compare both schemes in LPT to peculiar velocities in the Zeldovich Universe, noting that the recursive scheme is exact in this case. The left and right panels of Figure~\ref{fig:lpt_ir_resum} show the difference between these predictions for $v_s(\bk)$ and $\sigma^2_s(\bk)$, respectively, when the unresummed pieces are kept to 1-loop order. While the \textbf{C21} predictions do not diverge like in unresummed EPT at 1-loop, they are nonetheless substantially different than the Zeldovich Universe ``truth'' for $\sigma^2_s$, though they are essentially coincident for $v_s$ since the modifications to Equation~\ref{eqn:vw_window} from \textbf{C21} are at 1-loop order only. We can also compare the Zeldovich universe peculiar velocities to LPT predictions with a UV-IR split. Such predictions are shown in green for $k_{\rm IR} = 0.125 h$ Mpc$^{-1}$; while they differ from the full Zeldovich prediction due to lacking the contributions from short (UV) displacements at higher orders, they are in good perturbative agreement with the fully nonlinear prediction at large scales.

\section{Comparison to N-body Simulations}
\label{sec:nbody}

\subsection{Data}

In order to test the theoretical predictions in the previous sections we make use of the AbacusSummit simulations suite \cite{Garrison21,Maksimova21}.  In particular, we use \textsc{CompaSO} halo catalogs \cite{Hadzhiyska21} from the base set of simulations labeled \texttt{AbacusSummit\_base\_c000\_ph\{000-024\}}: these are $25$ boxes, each with volume $(2\  h^{-1} \text{Gpc})^3$ and $6912$ particles per side, run in a flat $\Lambda$CDM cosmology\footnote{Specifically, $\Omega_b h^2 = 0.02237$, $\Omega_c h^2 = 0.12$, $h = 0.6736$, $n_s = 0.9649$, $A_s = 2.083 \times 10^{-9}$ with one neutrino species with the minimal mass $m_\nu = 0.06 \text{eV}$ modeled as a non-clustering component.} derived from Planck satellite data \cite{PlanckCosmo18}. In order to explore the applicability of EFT predictions as a function of redshift, we consider a halo sample with masses  $\log_{10} \left(M/(h^{-1} M_\odot) \right) \in (12.5, 13.0)$ at redshifts $z=0.1, 1.1$.

We measure the \edit{redshift-space} density and velocity spectra of these halos in bins of width $dk = 0.005 h$ Mpc$^{-1}$ using \texttt{nbodykit} \cite{nbodykit}, painting the halos and their velocities onto $512^3$ grids using a triangular shaped cloud (TSC) scheme with interlacing. For all of our analyses we will use the mean measurements of the pairwise velocity moments of the $25$ boxes. For the purposes of fitting the data we also compute approximate covariance matrices in the disconnected (Gaussian) approximation based on the measured cross power spectra multipoles of the galaxy density, momentum and kinetic energy fields. These Gaussian covariances are in excellent agreement with the diagonal components of the numerical covariance matrix obtained from the $25$ AbacusSummit boxes. However, we caution that our fits using these Gaussian covariance matrices should be taken somewhat qualitatively given that galaxy density and peculiar velocity statistics become highly correlated at large scales, and also given that connected contributions to peculiar velocity statistics are known to be significantly enhanced relative to those in density statistics \cite{Howlett19}. Further details on computing the connected component of the covariance matrix are given in Appendix~\ref{app:gaussian_covariance}.

\subsection{Fits at Fixed Cosmology}

\begin{figure}
    \centering
    \includegraphics[width=\textwidth]{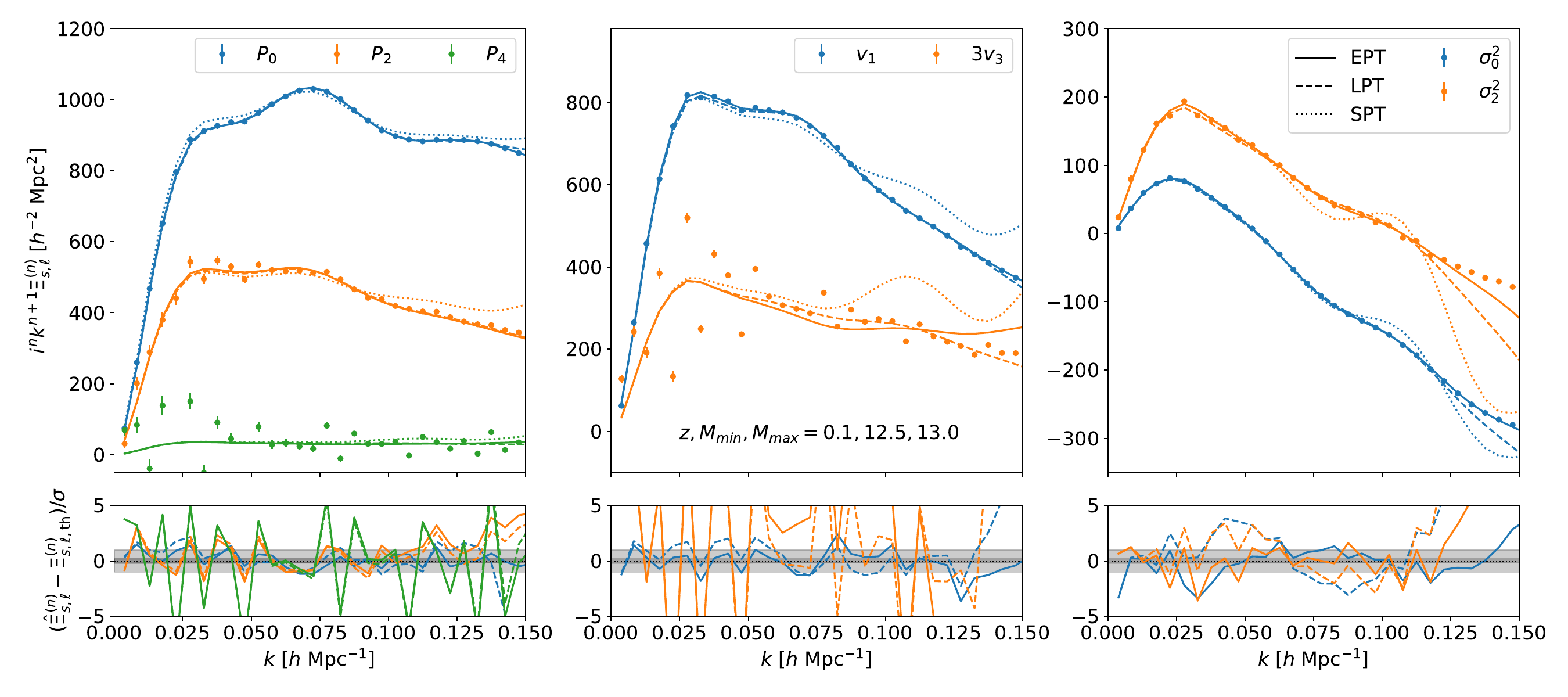}
    \includegraphics[width=\textwidth]{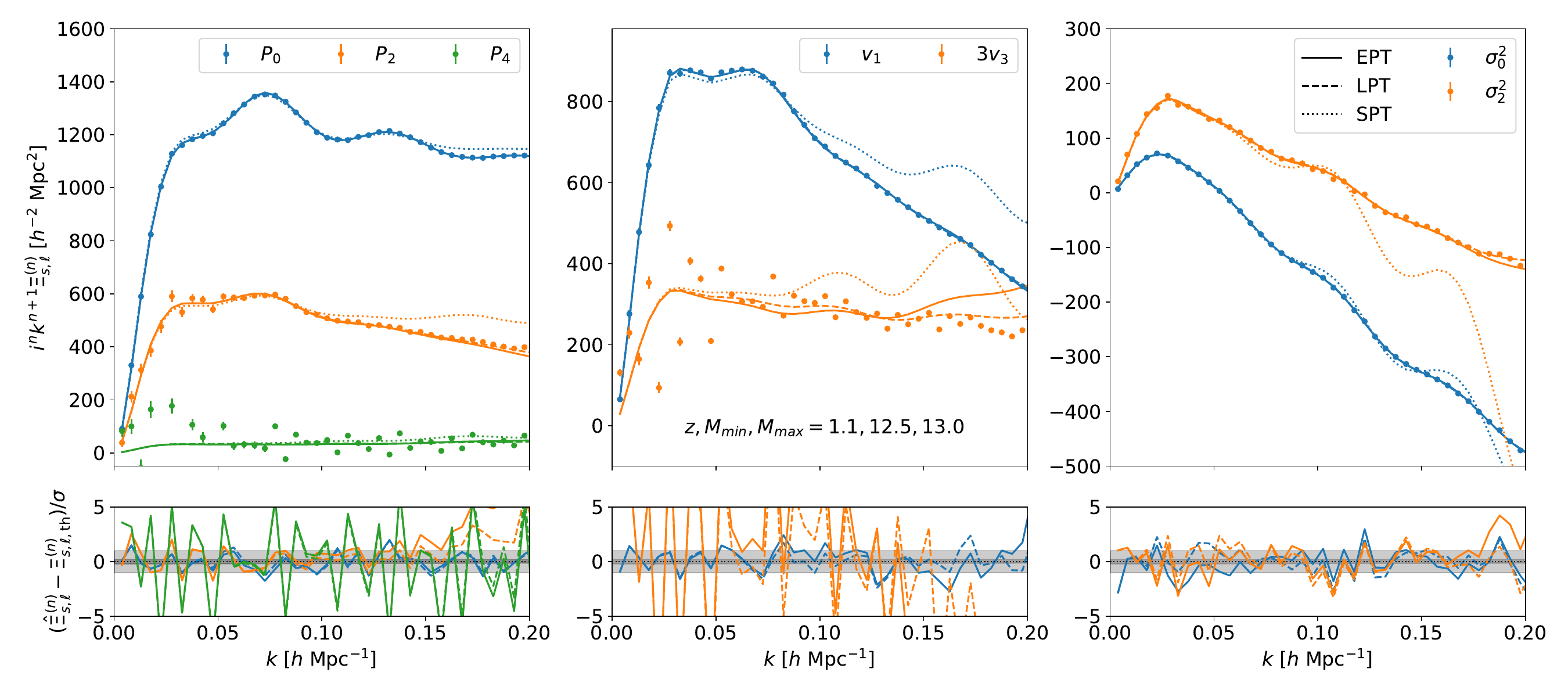}
    \caption{Fits to \edit{multipoles $\ell$ of} the first three redshift-space pairwise velocity moments for our two halo samples at fixed cosmology. Both EPT (solid) and LPT (dashed) are in very good agreement with the N-body measurements, with a better range of fit to smaller scales at the higher redshift, though we note that the EPT and LPT seem to better predict $\sigma_2^{s,2}$ and $v_{s,3}$, respectively. \edit{For comparison, dotted lines show un-resummed SPT predictions without EFT corrections.} The bottom panels shows residuals relative to error bars $\sigma$ for the mean of the $25$ boxes, each with volume $(2 h^{-1} \text{Gpc})^3$ box; the shaded gray band marks the $1\sigma$ region of the measurements. We note that discreteness effects in several spectra, particular in the higher multipoles, are significantly larger than the nominal statistical errors. }
    \label{fig:fit_logM_12.5_13.0}
\end{figure}

Figure~\ref{fig:fit_logM_12.5_13.0} shows fits to the first three \edit{redshift-space} pairwise velocity moments $P_s, v_s, \sigma^2_s$ in our fiducial mass range at redshifts $z=0.1$ and $z=1.1$. These fits were performed using only the auto-correlations of each spectrum rather than the full (disconnected) covariance---we will return to fits using the full covariance in the next subsection. Qualitatively we find that both EPT and LPT are able to produce good fits to the data, though \edit{there} is some evidence that the LPT fits perform slightly worse at the lower redshift $z = 0.1$, sacrificing the goodness of fit at large scales to compensate at higher $k$ when fit over the same range of scales. Unsurprisingly, both EFTs are able to better fit the data at higher redshift --- we find a qualitatively good fit for the halos at $z = 1.1$ out to $k_{\rm max} = 0.20\ h \text{Mpc}^{-1}$, with higher multipoles ($P_{s,2,4}, v_{s,3}, \sigma^2_{s,2}$) fit out to a slightly smaller $k_{\rm max} = 0.15\ h \text{Mpc}^{-1}$, compared to best-fit ranges of $k_{\rm max} = 0.15\ h \text{Mpc}^{-1}$ and $0.125\ h \text{Mpc}^{-1}$ for halos at $z = 0.1$. Also notable is that LPT slightly better fits the octopole of the first moment $v_{s,3}$, while EPT better matches the quadrupole of the second moment $\sigma^2_{s,2}$: the latter phenomenon was observed already in real space in ref.~\cite{Chen20}, while the former is quite noticeable at both redshifts shown. \edit{Since the EPT and LPT predictions shown differ only at 2-loop order or higher, the difference between these two formalisms should be taken as an indicator of a larger theoretical uncertainty for these two statistics.} However, we caution that the discreteness effects in $v_{s,3}$, as well as other higher multipoles like the power spectrum hexadecapole $P_{s,4}$ and to some extent $\sigma^2_{s,2}$, are significantly larger than the statistical error for the mean of the 25 boxes shown, which complicates a more detailed interpretation of goodness of fit for these statistics. This issue can be corrected for by accounting for discreteness effects in our theory predictions; however, since the signal associated with the most-affected spectra is rather weak, we will not pursue this avenue here and instead simply remove these spectra from our fits with the full covariance below.

\edit{
Making contact with the existing literature, the dotted lines in Figure~\ref{fig:fit_logM_12.5_13.0} also show fits to our N-body data using un-IR-resummed perturbation models without EFT corrections. We base these predictions on the standard perturbation theory (SPT) calculations of ref.~\cite{Okumura14}, which were extended to include cubic bias in ref.~\cite{Howlett19}, with a few major differences: firstly, as described in footnote 4 and Appendix~\ref{app:okumura}, these earlier papers feature a mis-organization of the perturbative expansion that lead to incorrect predictions with large infrared non-cancellations which we have corrected in this work. Secondly, these earlier works used an empirical extension of SPT where the velocity dispersions between galaxies that contribute to $P_{LL'}$ are treated as free parameters, leading to 2 additional independent parameters up to second order in the velocities; based on the discussion around Equation~\ref{eqn:fog_parameters}, we instead define FoG parameters $A_{\rm FoG}, B_{\rm FoG}$ such that
\begin{equation}
    P_{s,\rm lin}(k,\mu) \rightarrow (b + \lambda f \mu^2)^2 \left(1 - \frac12 \lambda^2 f^2 (k\mu)^2 A_{\rm FoG} + \frac{1}{24} \lambda^4 f^4 (k\mu)^4 B_{\rm FoG} \right),
\end{equation}
parameterizing the contributions from small-scale contributions to pairwise velocities in terms of their generating functions.\footnote{Specifically, stochastic contributions to the pairwise velocity $\Delta_\epsilon$ un-correlated with large-scale modes will contribute as a transfer function
\begin{equation}
    W_{\rm FoG}(k_\parallel) = \avg{e^{i k_\parallel (\lambda \Delta_{\epsilon,\parallel})}} = 1 - \frac12 \lambda^2 f^2 (k\mu)^2 A_{\rm FoG} + \frac{1}{24} \lambda^4 f^4 (k\mu)^4 B_{\rm FoG} + \ldots
\end{equation}
where $W_{\rm FoG}$ is given by the characteristic function of $\Delta_{\epsilon,\parallel}$ and $A_{\rm FoG}$ and $B_{\rm FoG}$ describe the first two nonzero moments of its probability distribution.
}
Finally, while previous works varied only the linear and quadratic biases $c_{\delta,\delta^2}$, fixing $c_{s^2,O_3}$ we allow all the dimensionless bias parameters at 1-loop---also employed in our EFT models---to vary independently in our fits along with constant corrections to the scale-independent shot noise. Despite these modifications, we find that the SPT models show a significantly reduced range of fit, only broadly matching the data when $k_{\rm max} < 0.10\ h$ Mpc$^{-1}$ at both redshifts, and even then sacrificing the agreement on large scales (e.g. the linear bias) for the overall goodness of fit at a level detectable at high significance given our simulations volumes. In addition, the lack of IR resummation leads to large deviations in the BAO amplitudes at small scales, with spurious enhancements in the BAO wiggles cleary visible by eye in Figure~\ref{fig:fit_logM_12.5_13.0}.
}

\begin{figure}
    \centering
    \includegraphics[width=\textwidth]{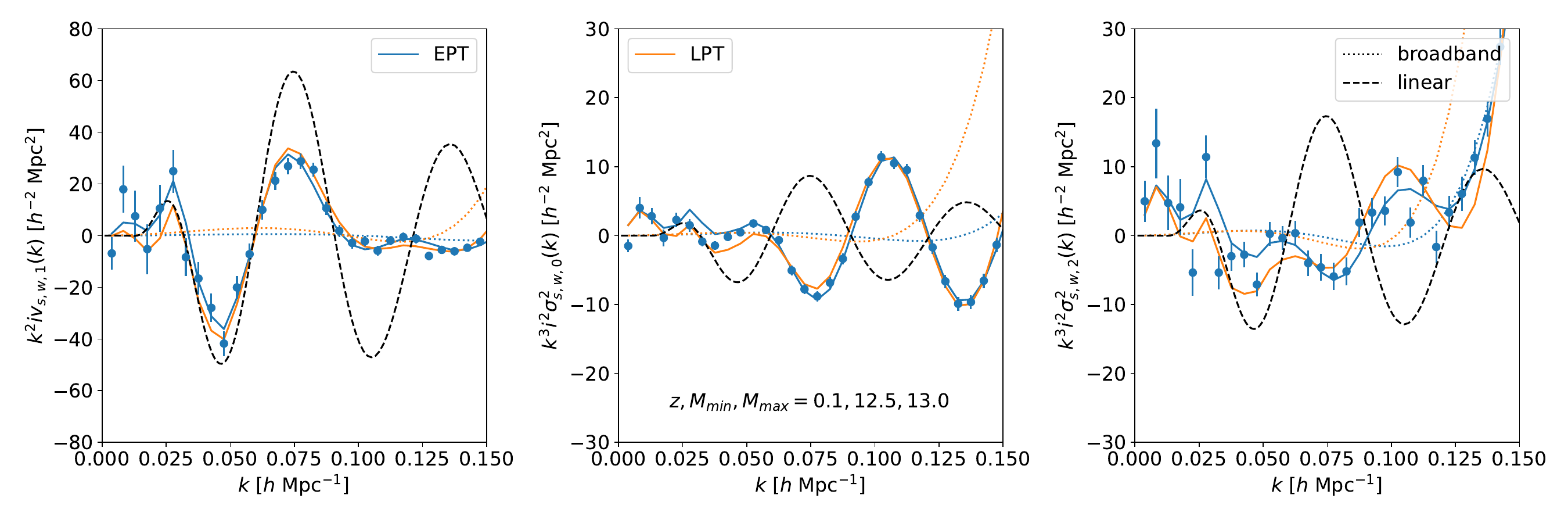}
    \includegraphics[width=\textwidth]{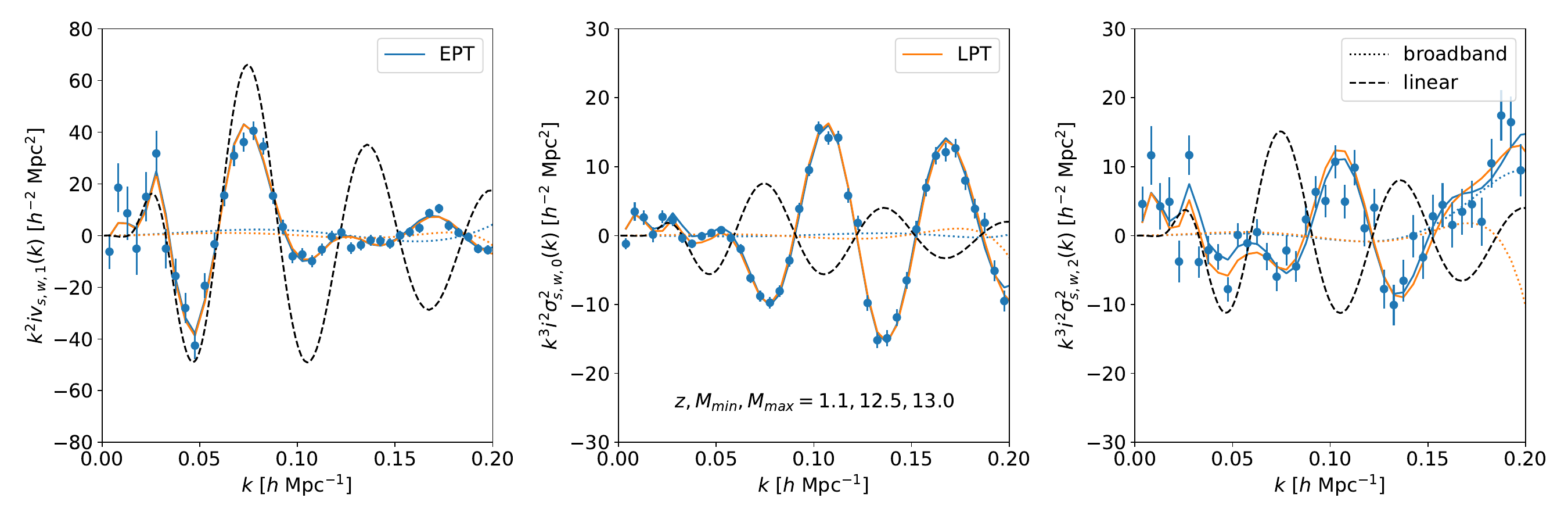}
    \caption{Predictions for the BAO feauture in \edit{redshift-space} peculiar velocity spectra for our two fiducial halo samples, computed by subtracting the contributions due to $P_{\rm lin}^{\rm nw}$ and, where needed, a broadband polynomial (dotted) forced to zero at large scales to account for differences from fitting the broadband power. Both the Eulerian and Lagrangian theories are in excellent agreement with the data even when the BAO signal is completely out of phase with linear theory (dashed). }
    \label{fig:wiggles_logM_12.5_13.0}
\end{figure}

In order to investigate the effect of IR-resummation on the BAO, in Figure~\ref{fig:wiggles_logM_12.5_13.0} we subtract off the broadband prediction computed using $P_{\rm L}^{nw}$ and the best-fit EFT parameters. In order to remove any additional broadband effects, we  include a quartic polynomial in $k^2$ fit to the difference between the theory and data, with the constant term set to zero to force the low $k$ prediction to agree with linear theory. Both EPT and LPT (blue and orange) are able to predict the envelope of the BAO wiggles, with accuracy comparable to the statistical precision of the measurements, even when the amplitude and phase of the BAO signal are significantly modified compared to the linear theory prediction (black dashed). The small disagreements are more pronounced at lower redshifts, where the underlying nonlinearities (including in the broadband) are less well-modeled by perturbation theory, and particularly in the quadrupole $\sigma^2_{s,2}$ where discreteness effects at low $k$ are significant compared to the amplitude of the BAO signal. Nonetheless, even for the $z=0.1$ second-moment quadrupole $\sigma^2_{s,2}$ it is apparent that the linear-theory prediction is in substantial tension with the data, while both EPT and LPT agree within the statistical and systematic scatter of the measurements. The ability of our IR-resummation schemes to broadly capture the shape of the BAO in these peculiar velocity statistics represents a new validation of the theory of the effect of long-wavelength bulk motions on the BAO in an arena beyond typical spectroscopic galaxy surveys.

\subsection{Fits with Free Growth Rates}

\begin{figure}
    \centering
    \includegraphics[width=0.49\textwidth]{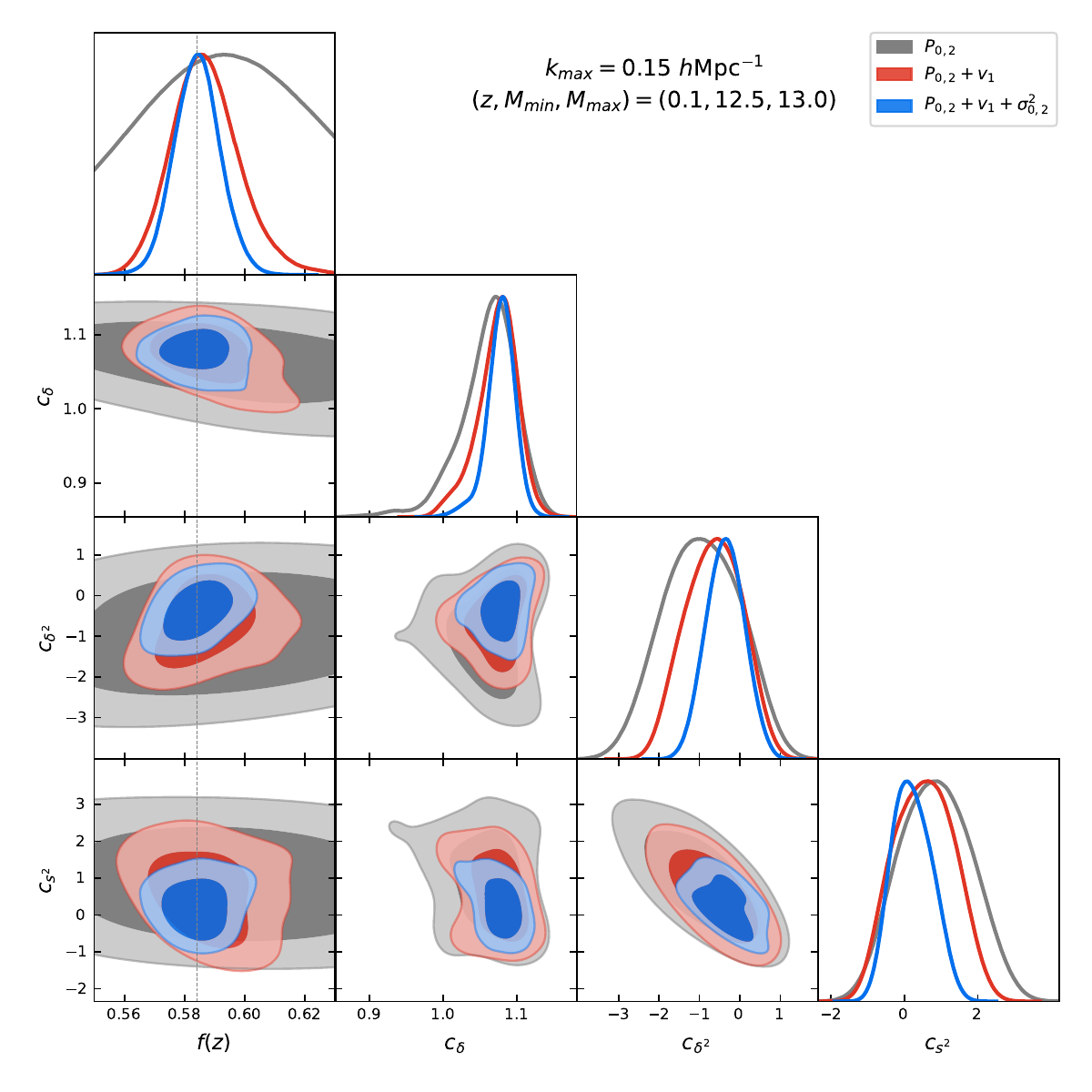}
        \includegraphics[width=0.49\textwidth]{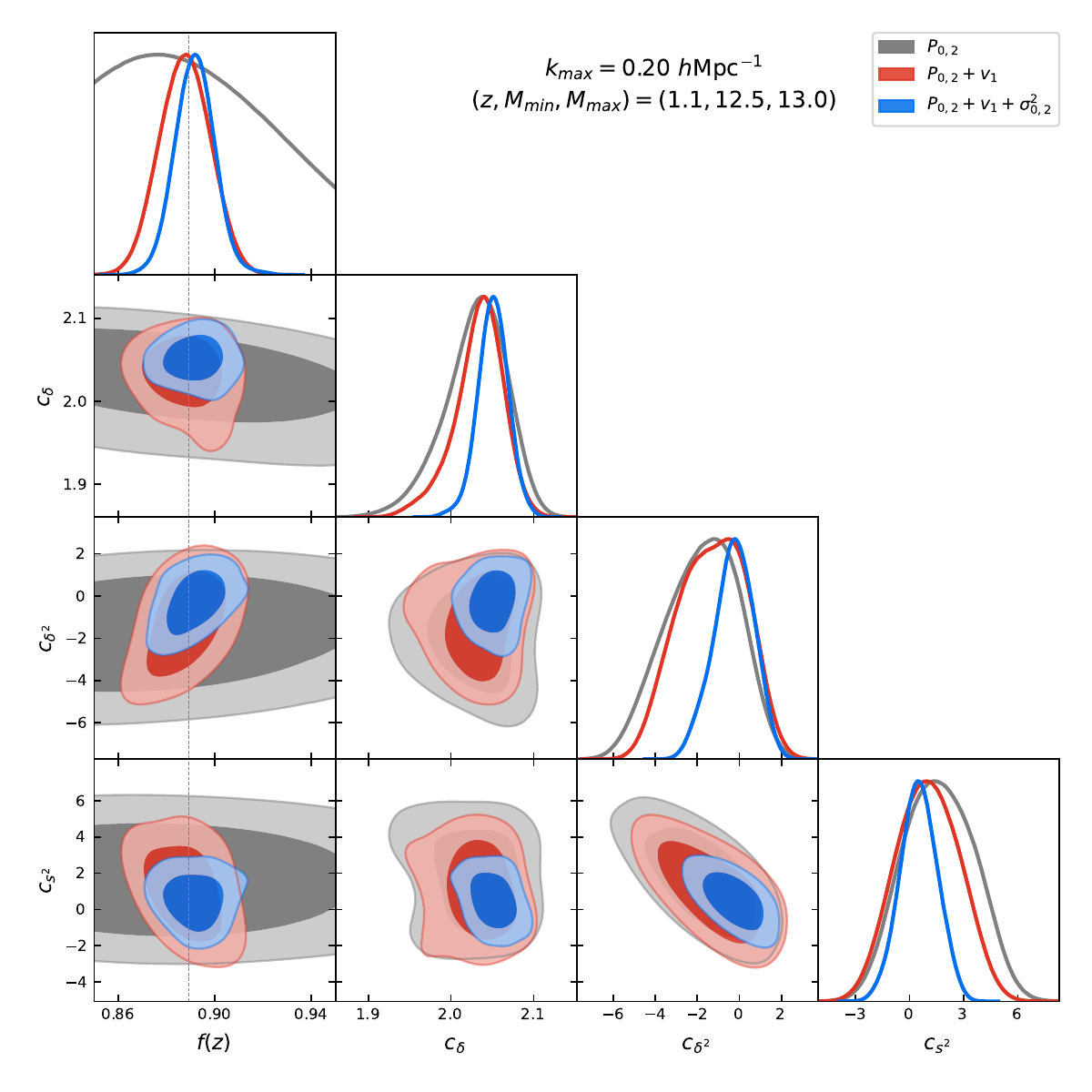}
    \caption{Constraints on the growth rates $f(z)$ using the galaxy power spectrum and first two pairwise velocity moments, using the two fiducial halo samples at $z= 0.1, 1.1$ (left and right). All statistics are fit with the same set of free EFT parameters with wide, uninformative priors---adding in successively higher moments of the pairwise velocity breaks degeneracies and cancels cosmic variances, leading to a significant ($> 5\times$) improvement in the constraints on $f(z)$, which remain unbiased in all cases to better than half a percent.}
    \label{fig:fz_fits}
\end{figure}

One of the most significant promises of galaxy peculiar velocity measurements is the ability to enhance measurements of the growth of structure at low redshifts from spectroscopic galaxy surveys. While the application of our perturbative models to realistic survey data is beyond the scope of this work, we can conduct a preliminary investigation using the halo peculiar velocity spectra above and their full disconnected covariances. In order to do so, we will fit the data keeping all cosmological parameters fixed but allow the linear growth rate $f(z)$ to vary as a free parameter while also fitting for all the EFT parameters. Unlike in the previous subsection where we used only the purely-diagonal components of the covariance matrix, in this subsection we will use the full disconnected covariance for 1 box only --- we make this choice in order to account for the covariance of galaxy densities and velocities while not over-weighting residual systematics such as discreteness effects or higher-loop contributions \cite{Maus25b}. The disconnected covariance matrix of the pairwise velocity statistics is very close to singular at large scales due to the high level of correlation between the densities and velocities at low $k$, where both fields effectively sample the linear initial conditions. Since the goal of this work is to develop the modeling of peculiar velocity statistics in the mildly nonlinear regime, rather than incorporating further realism on large scales into our covariances such as velocity errors and connected contributions which would break these degeneracies, we simply apply a cut at $k_{\rm min} = 0.04$ ($z=0.1$) and $0.035\ h \text{Mpc}^{-1}$ ($z = 1.1$) where the disconnected covariance becomes well behaved (Appendix~\ref{app:gaussian_covariance}). We also drop the power spectrum hexadecapole $P_4$ and first-moment octopole $v_3$ --- neither of which are expected to significantly contribute to cosmological constraints --- due to their significant discreteness effects. We restrict our attention to fits using the Eulerian EFT in this exploratory section since predictions in the Lagrangian EFT take significantly longer to compute; we leave applying the latter to realistic data, for example via a Taylor series expansion in cosmological parameters, to future work.

Figure~\ref{fig:fz_fits} shows the one- and two-dimensional posteriors for $f(z)$ and the dimensionless bias parameters when fitting to the power spectrum and first two pairwise velocity moments of the two redshift samples considered in this work. While it would be a somewhat suboptimal practice for realistic data analyses, in these fits we have adopted wide, uninformative priors on the EFT parameters, using the same set of free parameters as given in Equations~\ref{eqn:bias_parameters_ept}, \ref{eqn:counterterms}, \ref{eqn:stoch} and \ref{eqn:fog_parameters} regardless of the data combination used. By doing so, we can see that adding successively higher-order pairwise velocity statistics to the fit allows us to obtain tighter constraints through a combination of sample variance cancellation and degeneracy breaking. Indeed, adding up to the second moment decreases the error bar on $f(z)$ by more than a factor of $5$ at either redshift, yielding $1.3\%$ and $0.9\%$ constraints on the growth rate at $z = 0.1$ and $z = 1.1$, respectively. \edit{We caution that this factor of five improvement shouldn't be taken as a forecast for realistic data settings, since observational systematics and velocity errors will necessarily change the weighting of information content between the different statistics; rather, our joint fits demonstrate that sufficiently many nonlinear shapes in velocity spectra exist in principle to break the degeneracies between the full set of EFT parameters at 1-loop order. We refer the interested reader to e.g. refs.~\cite{Koda2014,Sugiyama17,Howlett2017a,Rosselli2025} for cosmological forecasts including peculiar velocities in more realistic settings.

Last but not least, the constraints shown in Figure~\ref{fig:fz_fits} are unbiased,} with best-fit values deviating at most $0.3\%$ from the truth.\footnote{We note that this level of accuracy is tighter than the expected error due to the scale-independent neutrino approximation used in AbacusSummit \cite{Maksimova21}, though our dropping of the largest scales due to difficulties in covariance matrix estimation also removes the scales most sensitive to this lack of scale dependence.} Even if we naively scale the covariance down to account for the volume of all $25$ boxes, these best fit values are still less than $2\sigma$ away from the truth given the reduced covariance. The resulting best-fit curves for each velocity spectrum are shown in Appendix~\ref{app:best_fit_curves} --- the corresponding residuals comparing to the measured spectra are smaller than the statistical errors of a $8 h^{-3}$ Gpc$^3$ box, which is in turn significantly larger than any expected peculiar velocity survey. We note, however, that the fit is less good at the level of the reduced $25$-box ($200 h^{-3}$ Gpc$^3$) covariance, with large contributions to $\chi^2$ from the large-scale second-moment quadrupole $\sigma^2_2$ that are likely due to discreteness effects. Nonetheless, especially given that we are likely underestimating the errors on these measurements by using an approximate, disconnected covariance, our fits show that EFT predictions are more than ready for the coming era of peculiar velocity surveys.

\section{Conclusions}
\label{sec:conclusions}

Galaxy peculiar velocities, inferred via redshift-independent measures of distances like the fundamental plane, the Tully-Fisher relation, or supernovae observations, offer a direct probe of the cosmological growth of structure, particularly at late times where complementary constraints of the growth rate $f\sigma_8$ from spectroscopic galaxy surveys are limited by cosmic variance. In the coming decade, surveys like DESI and the Rubin observatory, among others, will provide unprecedented peculiar velocity measurements at low redshifts, allowing for stringent tests of structure formation where the influence of dark energy is most pronounced. Since these new surveys will probe structure formation at its most nonlinear epoch, careful modeling of the underlying gravitational nonlinearities will be especially important.

In this paper, we have extended calculations in the effective field theory (EFT) of large-scale structure (LSS) developed in the context of redshift-space galaxy clustering and spectroscopic galaxy surveys --- for which we give a brief review in Section~\ref{sec:pt_overview} --- to study galaxy peculiar velocity statistics. We do so using both the Eulerian and Lagrangian formulation of the EFT, which describe structure formation via densities and velocities at fixed positions, or the trajectories of infinitesimal fluid elements, respectively. In particular, we have focused on predicting pairwise velocity statistics, specifically the Fourier transforms of the density-weighted moments of the velocity difference $\bv_{p,1} - \bv_{p,2}$ between pairs of galaxies as a function of their separation. Since they are the difference between pairs of velocities, pairwise velocity statistics are Galilean invariant and insensitive to large-scale bulk flows that may otherwise affect peculiar velocity measurements. These statistics also have the convenient feature that their generating function is the galaxy redshift-space power spectrum \cite{Sugiyama16}, so that for any sample of galaxies they can be simultaneously modeled with the galaxy density power spectrum with one consistent set of parameters, which we describe in Section~\ref{sec:pv_pt}.

The relation between galaxy density and velocity spectra also has interesting implications for their infrared (IR) resummation. The presence of the baryon acoustic oscillations (BAO) feature introduces a new scale and numerically-large parameter ($k \Sigma_d \gtrsim 1$) into the nonlinear clustering of galaxies; this large parameter, due to large-scale bulk flows and velocities, has to be resummed in order to keep the perturbation theory well-behaved, resulting in an exponential damping of the oscillatory BAO signal. However, as we show in Section~\ref{sec:pv_pt}, since peculiar velocity statistics directly probe the bulk flows causing this effect, they require a modified treatment, such that both the sign and amplitude of the BAO in peculiar velocity statistics can be dramatically different from that in galaxies. In Section~\ref{sec:pv_zel}, we explore these effects in a toy Zeldovich universe where peculiar velocity statistics can be computed to all orders, showing that our updated IR resummation scheme is able to accurately describe the BAO signal therein, whereas the un-resummed theory or the naive application of earlier schemes developed for galaxy densities lead to uncontrolled behavior.

Our calculations in this paper, as described above, set the stage for future analyses of peculiar velocity data within a consistent EFT framework. As a proof of principle, in Section~\ref{sec:nbody} we measure the peculiar velocity statistics of halos in the \texttt{AbacusSummit} simulations and compare them to theoretical predictions from the EFT. Our analytic predictions in both the Eulerian and Lagrangian EFTs well-describe the fully nonlinear measurements from N-body simulations on perturbative scales, with a wider range of fit when going to higher, more linear redshifts. We further isolate the BAO feature in these measurements and find them to be in excellent agreement with the theory of IR resummation. Looking forward to more realistic analyses, we conduct fits to these N-body data varying the growth rate $f(z)$, finding that our models are able to recover the growth rate even at low redshifts at below the percent level, well beyond the statistical precision required for upcoming peculiar velocity surveys. We make our \texttt{Python} code for these calculations, \texttt{velocisaurus}, publicly available.\footnote{\url{https://github.com/sfschen/velocisaurus}}

Let us close by noting some possible avenues for future work. On the theoretical side, while we have focused on computing galaxy pairwise velocity statistics, it should be straightforward to extend the EFT calculations in this work to other commonly measured velocity statistics such as the auto-spectrum of the galaxy momentum field $\rho = (1 + \delta) v$ \cite{Okumura14,Dam21}, or to analyze galaxy velocities at the field level \cite{Lai23,Ravoux25}. Further afield, peculiar velocities can also be measured from the kinetic Sunyaev-Zeldovich effect in the cosmic microwave background (see e.g.~ref.~\cite{Smith18}), whose statistics have yet to be predicted within the EFT.  Finally, on the more pragmatic side, while we have performed some preliminary steps towards analyzing peculiar velocity data derived from halos in N-body simulations, an obvious next step will be to apply our modeling to realistic simulated mocks with non-Gaussian covariance matrices, in expectation for the eventual goal of analyzing upcoming data. While this paper was dedicated to better modeling the onset of nonlinearity towards small scales, since peculiar velocity statistics derive significant amounts of statistical power on large scales it will be equally important to include both general-relativistic effects and wide angle effects present in realistic mocks into predictions for future surveys \cite{Castorina19,Pantiri24}. At the low redshifts relevant for peculiar velocity surveys, how to gracefully transition between predictions in these two regimes will be a question of practical importance.

\section*{Acknowledgements}

We thank Teppei Okumura, Zvonimir Vlah, Martin White, Nickolas Kokron and Matias Zaldarriaga for useful discussions.

SC acknowledges support from the National Science Foundation at the IAS through NSF/PHY 2207583. Support for this work was provided
by NASA through the NASA Hubble Fellowship grant
HST-HF2-51572.001 awarded by the Space Telescope
Science Institute, which is operated by the Association of
Universities for Research in Astronomy, Inc., for NASA,
under contract NAS5-26555. This work was performed in part at the Aspen Center for Physics, which is supported by National Science Foundation grant PHY-2210452. YL and CH acknowledge support from the Australian Government through the Australian Research Council’s Laureate Fellowship (project FL180100168) and Discovery Project (project DP20220101395) funding schemes. YL is also supported by an Australian Government Research Training Program Scholarship. FQ acknowledges support from the Excellence Initiative of Aix-Marseille University - A*MIDEX, a French ``Investissements d'Avenir'' program (AMX-20-CE-02-DARKUNI).

\appendix

\section{Angular Integrals in Lagrangian Perturbation Theory}
\label{app:lpt_integrals}

In this Appendix we review the methods developed in ref.~\cite{Chen21} to compute the angular integrals required to recast integrals of the form Equation~\ref{eqn:clpt} into Hankel transforms, allowing them to be efficiently computed using the FFTLog algorithm \cite{fftlog}. Specifically, we will extend \textbf{Method II} in ref.~\cite{Chen21} to cover the integrals required for redshift-space velocity statistics.

Computing peculiar velocity statistics in LPT as in Equation~\ref{eqn:xin_lpt} generically leads to integrals of the form
\begin{equation}
    \Xi^{(n)}_s \supset \hn_{i_1} ... \hn_{i_n} \int d^3\bq\ e^{i\bk\cdot\bq - \frac12 k_i k_j A^{s, \rm lin}_{ij}} k_{j_1} ... k_{j_m} f^s_{i_1 ... i_n j_1 ... j_m}(\bq) \nonumber
\end{equation}
where the tensor function in the integrand is formed from products of expectation values involving $\dot{\Delta}$, $\Delta^{(n)}$ and scalar bias operators; schematically, these can look, for example, like
\begin{equation}
    f^s_{i_1 ... i_n j_1 ... j_m}(\bq) \sim \langle \dot{\Delta}^{(a_1)}_{i_1} \Delta^{s,(b_1)}_{j_1} O(\bq) \rangle \ldots \avg{\ldots} \nonumber
\end{equation}
and can be recast as expectation values in real space, e.g. 
\begin{equation}
    f^s_{i_1 ... i_n j_1 ... j_m}(\bq) \sim a_1 f(a)  R^{(b_1)}_{j_1 l_1} \langle \Delta^{(a_1)}_{i_1} \Delta^{(a_1)}_{l_1} O(\bq) \rangle \ldots \avg{\ldots}. \nonumber
\end{equation}
These real-space correlators can be decomposed by symmetry into tensor products of the unit vectors $\hn, \hq, \hk$ multiplied by scalar functions of $k, q$, such that their most general form is
\begin{equation}
    \int d^3\bq\ e^{i\bk\cdot\bq - \frac12 k_i k_j A^{s, \rm lin}_{ij}(\bq)} (\hn \cdot \hq)^n (\hk \cdot \hq)^m f(k,q).
    \label{eqn:lpt_form}
\end{equation}
Our task is thus to solve integrals in this form generally.

In \textbf{Method II}, we redefine
\begin{equation}
    k_i k_j A_{ij}^{s, \rm lin}(\bq) = K_i K_j A_{ij}^{\rm lin}(\bq), \quad K_i = R^{(1)}_{ij} k_j
\end{equation}
such that the quadratic exponent has no azimuthal dependence in the spherical coordinate system defined around $\textbf{K}$. In these coordinates, $\bq = q (\sqrt{1-\mu_{\bq}^2} \cos\phi, \sqrt{1-\mu_{\bq}^2}\sin\phi, \mu_{\bq})$, and we can write
\begin{align}
    &\hn \cdot \hq = A(\mu) \mu_{\bq} + B(\mu) \sqrt{1-\mu_{\bq}^2} \cos\phi, \nonumber \\
    &\hk \cdot \hq = c(\mu) \mu_{\bq} - s(\mu) \sqrt{1-\mu_{\bq}^2} \cos\phi
\end{align}
where we have defined in terms of the line-of-sight angle $\mu = \hn \cdot \hk$
\begin{align}
    &A(\mu) = \frac{(1+f)\mu}{\sqrt{1 + f(2+f)\mu^2}}, \quad B(\mu) = \sqrt{\frac{1-\mu^2}{1 + f(2+f)\mu^2}}, \quad c(\mu) = \frac{1 + f\mu^2}{\sqrt{1 + f(2+f)\mu^2}},
\end{align}
and $s(\mu) = \sqrt{1 - c^2(\mu)}$. The above coordinates allow us rewrite Equation~\ref{eqn:lpt_form} in the form
\begin{equation}
    I_{n,m}(A,B,C) = i^{m-n} \int d \mu_{\bq} d \phi\ e^{-iC\sqrt{1-\mu_{\bq}^2} \cos\phi + i A \mu_{\bq} + B \mu_{\bq}^2} \left( \sqrt{1-\mu_{\bq}^2} \cos\phi \right)^n \mu_{\bq}^m
    \label{eqn:mii_general}
\end{equation}
where the first two terms in the exponent come from writing the Fourier transform term $\bk \cdot \bq$ in $\textbf{K}$ coordinates. 

As shown in ref.'s~\cite{Vlah19,Chen21}, Equation~\ref{eqn:mii_general} can be solved as an infinite series in spherical Bessel functions
\begin{equation}
    I_{n,m}(A,B,C) = 4\pi e^B \sum_{\ell=0}^\infty \left( \frac{-2}{\rho} \right)^\ell \tilde{G}^{(m)}_{n,\ell}(A,B,\rho) j_\ell(\rho)
\end{equation}
where $\rho = A^C + C^2$. Since $\rho = kq$ in the specific example of~\ref{eqn:lpt_form}, the remaining radial integral in $q$ is now a Hankel transform and can now be computed via FFTLog. Importantly, the coefficients $\tilde{G}^{(m)}_{n,\ell}$ obey the recursion relations $\tilde{G}^{(m)}_{n,\ell} = \partial_A \tilde{G}^{(m-1)}_{n,\ell} + A \tilde{G}^{(m)}_{n,\ell-1}/2$ and $\tilde{G}^{(m)}_{n,\ell} = \partial_C \tilde{G}^{(m)}_{n-1,\ell} + C \tilde{G}^{(m)}_{n,\ell-1}/2$, with the first element of the series given by\footnote{These expressions correct minor typos in Appendix C of ref.~\cite{Chen21}.}
\begin{equation}
    \tilde{G}^{(0)}_{0,\ell} = \sum_{a=\ell}^\infty f_{a\ell} \left( \frac{BA}{\rho^2} \right)^a {}_2 F_1(\frac12 - a, -a; \frac12 - a - \ell;  \frac{\rho^2}{A^2}).
\end{equation}
Here we have defined
\begin{equation}
    f_{a\ell} = \frac{\Gamma(a+\ell+\frac12)}{\Gamma(a+\frac12)\Gamma(\ell+1)\Gamma(1-\ell+a)}
\end{equation}
and ${}_2F_1$ is the ordinary hypergeometry function.

The above arguments therefore show that redshift-space peculiar velocity statistics in LPT can be computed using the same numerical methods developed for the redshift-space power spectrum, ie. by recasting each contribution into the form of Equation~\ref{eqn:mii_general}. The only significant difference is that, due to the tensor structure of the peculiar velocities, their integrals require that $I_{n,m}$ be evaluated to higher indices, requiring higher derivatives of $\tilde{G}^{(0)}_{0,\ell}$. Each of these can be written in the form
\begin{align}
    \frac{\partial^{s+t} \tilde{G}^{(0)}_{0,\ell} }{\partial A^s \partial C^t} = \sum_{a=\ell}^\infty f_{a\ell} \left( \frac{BA}{\rho^2} \right)^a &\Big( P_{st}(A,C,a,\ell) {}_2F_1(\frac12 - a, -a; \frac12 - a - \ell;  \frac{\rho^2}{A^2}) \nonumber \\
    &+ Q_{st}(A,C,a,\ell) {}_2F_1(\frac32 - a, -a; \frac12 - a - \ell;  \frac{\rho^2}{A^2}) \Big);
\end{align}
we refer the reader to Appendix C of ref.~\cite{Chen21} for the lower-order derivatives and list the new ones below.

\begin{align}
    P_{03} &= - \frac{1}{s\rho^3} \left(2 (2 a-1) c^4 (l+1) (2 a-2 l-1)+c^2 s^2 (-28 a (l+1)+14 l+5)+6
   s^4\right) \nonumber \\
   Q_{03} &= - \frac{1}{s\rho^3} (2 a-1) \left(2 c^4 (2 a-2 l-1) (a-l-1)+c^2 s^2 (-4 a (l+4)+14 l+5)+6
   s^4\right) \nonumber \\
   P_{12} &= \frac{1}{c\rho^3} \left(2 c^4 \left(4 a^2 (l+1)-4 a l (l+1)+2 l^2+l\right)-c^2 s^2 (20 a (l+1)-10
   l+1)+2 s^4\right) \nonumber \\
   Q_{12} &= \frac{1}{c\rho^3} (2 a-1) \left(2 c^4 (2 a-2 l-1) (a-l)-c^2 s^2 (4 a (l+3)-10
   l+1)+2 s^4\right) \nonumber \\
   P_{22} &= - \frac{1}{\rho^4} \big(-c^2 s^2 \left(16 a^2 (l+1) (l+5)-80 a \left(l^2-1\right)+12 l (3
   l-2)+3\right)\nonumber \\
   &+2 c^4 \big(8 a^3 (l+1)  -4 a^2 (l+1) (4 l-1)+2 a (l+1) \left(4
   l^2+1\right)-4 l^3+l\big) \nonumber \\
   &+12 s^4 (4 a (l+1)-2 l+1)\big) \nonumber \\
   Q_{22} &= -\frac{1}{\rho^4} (2 a-1) \big(-c^2 s^2 \big(4 a^2 (4 l+11)-16 a (l (l+4)-2)+12 l (3
   l-2)+3\big) \nonumber \\
   &\quad +2 c^4 (2 a-2 l-1) (2 a-2 l+1) (a-l)+12 s^4 (2 (a-1) l+3 a+1)\big) \nonumber \\
   P_{31} &= \frac{s}{c\rho^4} \big(4
   c^4 \big(\big(4 a^3+a-1\big) l+2 (2 a-1) l^3+(3-2 a (4 a+1)) l^2+a (4 a
   (a+2)+7)\big) \nonumber \\
   &-4 c^2 s^2 \left(2 a^2 (l+1) (2 l+7)+a ((9-14 l) l+23)+6 l^2-9
   l+3\right) \nonumber \\
   &+3 s^4 (4 a (l+1)-2 l+1)\big) \nonumber \\
   Q_{31} &= \frac{s}{c\rho^4} (2 a-1) \big(-2 c^2 s^2 \big(8 a^2 (l+2)+a (25-2 l (4 l+7))+6 (l-1) (2
   l-1)\big) \nonumber \\
   &\quad +4 c^4 (2 a-2 l+1) (a-l) (a-l+1)+3 s^4 (4 a (l+1)-2 l+1)\big)
\end{align}

\section{Redshift-Space Power Spectrum Counterterms in LPT}
\label{app:counterterms}

In this appendix, we review te derivation of the counterterm structure of the redshift-space power spectrum in LPT, following the excellent discussions in refs.~\cite{Vlah15,Ebina24} while highlighting the velocity counting in $\lambda$. Since galaxy formation is a spatially local process, the bias expansion in Equation~\ref{eqn:bias_expansion} was expressed in terms of observables like the overdensity and tidal field evaluated at the coordinates of the galaxy. This locality breaks down at the halo scale $R_h$, and on large scales this effect can be described using a derivative bias
\begin{equation}
    F_g(\bq) \supset b_{\nabla^2} \nabla^2 \delta_0(\bq) + \epsilon(\bq)
\end{equation}
where $b_{\nabla^2} \sim R_h^2$ and in addition we have included a contribution $\epsilon$ to account for small-scale stochasticity with only local correlations (i.e. $\avg{\epsilon(\bq) \epsilon(\textbf{0})} = 0$ if $q \gg R_h$). Similarly the displacement receives corrections
\begin{equation}
    \Psi_i \supset c_0 \nabla_i \delta_0(\bq) + \epsilon_i(\bq).
\end{equation}
Since we are also interested in velocities we will need in addition that $\dot{\Psi} \supset \dot{c_0} \nabla_i \delta_0 + \dot{\epsilon}_i$. The velocity contributions are independent since we only look at the time at which the galaxies are observed.

Finally, in LPT and in dealing with peculiar velocity statistics we will frequently have to deal with products of observables at a single point, e.g. the velocity squared $\bu_{\hn}^2$. These observables have their own small-scale dependences, for example at 1-loop order \cite{Porto14,Vlah15,Ebina24}
\begin{align}
    [\Psi_i \Psi_j](\bq) &= (\Psi_i \Psi_j)^{\rm PT} + \beta_1 \delta_{ij} + \beta_2 \delta_{0}(\bq) \delta_{ij} + \beta_3 s_{0,ij}(\bq) + \epsilon_{ij}(\bq) \nonumber \\
    [\delta \Psi_i)](\bq) &= (\delta \Psi_i)^{\rm PT} + \beta_4 \nabla_i \delta_0(\bq) + \epsilon_i(\bq)
\end{align}
where the superscript PT denotes the parts of these operators given by perturbation theory mode coupling kernels and $\beta_n$ are constants. These extra degrees of freedom arise to capture the statistics of products of stochastic modes, such that e.g. the correlator $\avg{\epsilon_i(\bq) \epsilon_j(\textbf{0})}$ approaches the product operator averaging over stochastic modes, e.g. $\avg{\epsilon_i \epsilon_j} = \beta_1 \delta_{ij}$, in the limit $\bq \rightarrow \textbf{0}$.\footnote{For simplicity we have dropped the dependence of the short modes on long-wavelength perturbations which would be important when going to higher order.} Beyond the mean values $\beta_n$ we will also need to track the stochastic fluctuations induced by short-wavelength modes, denoted here by $\epsilon_{i}, \epsilon_{ij}$.  We will denote corrections to equivalent operators featuring velocities, ($\Psi \dot{\Psi}, \dot{\Psi}\dot{\Psi}$ etc.), using the same dotted notation as above ($\dot{\beta}, \dot{\epsilon}$, etc.) following ref.~\cite{Ebina24}. At 1-loop order we also encounter products of 3 and 4 operators (e.g. $\Psi\Psi\Psi\Psi$) but the corrections for these contact terms renormalize beyond 1-loop contributions which we will therefore not consider.

Combining the above terms into Equation~\ref{eqn:generating_function} at one-loop order gives the counterterm contributions \cite{Ebina24}
\begin{align}
    P_s(&\bk;\lambda) \supset 2\ \big(1 + b_\delta + \lambda f\mu^2 \big)\ \Big( b_{\nabla^2} - c_0 - \lambda \dot{c}_0 \mu^2 - \frac12 (\beta_1 + (2\lambda \dot{\beta}_1 + \lambda^2 \ddot{\beta}_1) \mu^2) (1 + b_\delta + \lambda f\mu^2)  \nonumber \\
    &+ \frac12 \big( \beta_2 + \frac23 \beta_3 + 2 \lambda (\dot{\beta}_2 + \frac23 \dot{\beta}_3) \mu^2 + \lambda^2 \mu^2 (\ddot{\beta}_2 + (\mu^2 - \frac13) \ddot{\beta}_3) \big) - b_\delta (\beta_4 + \lambda \dot{\beta}_4 \mu^2) \Big)  k^2 
 P_{\rm lin}(k). \nonumber
\end{align}
The coefficients to $\lambda^n$ above fully define the counterterm contributions to the $\tilde{\Xi}^{(n)}_s$. However, the above form is cumbersome and only some of the 14 free paramters are degenerate---in particular we can rewrite the above as
\begin{align}
    P_s(&\bk;\lambda) \supset 2\ \big(1 + b_\delta + \lambda f\mu^2 \big)\ k^2 \ \Big( \beta_0^0 + (\lambda \beta_2^1 + \lambda^2 \beta_2^2) \mu^2 + (\lambda^2 \beta_4^2 + \lambda^3 \beta_4^3) \mu^4 \Big) P_{\rm lin}(k)
\end{align}
from which we see there are only $5$ free parameters, and indeed only $3$ in the case of the redshift-space power spectrum ($\lambda=1$). As a consequence, only 5 counterterms to the pairwise velocity spectra at 1-loop are independent in both real and redshift space

\section{IR Resummation of Peculiar Velocities in Real Space}
\label{app:ir_resummation_real_space}

The IR resummation of the BAO for peculiar velocities in real space follows upon very similar lines to the discussion in Section~\ref{ssec:ir_resum_pv}---indeed we can carry out the identical derivation leading up to e.g. Equation~\ref{eqn:new_contribution_vk} by simply substituting redshift space quantities with subscripts `` $s$ '' with their real-space counterparts. However, a more straightforward approach is to simply use the derivative formulat (Eqn.~\ref{eqn:generating_function}) at $\lambda = 0$. In this case have
\begin{align}
    P_w(\bk) &= e^{-\frac12 k^2 \Sigma^2_d} \left( \llbracket P \rrbracket_w \right)(\bk) \nonumber \\
    v_w(\bk) &= e^{-\frac12 k^2 \Sigma^2_d} \left( \llbracket v \rrbracket_w + (ik\mu) f \Sigma^2_d \llbracket P \rrbracket_w \right)(\bk) \nonumber \\
    \sigma^2_w(\bk) &= e^{-\frac12 k^2 \Sigma^2_d} \left( \llbracket \sigma^2 \rrbracket_w + 2 (ik\mu) f \Sigma^2_d \llbracket v \rrbracket_w + (f^2 \Sigma_d^2 + (i k \mu)^2 f^2 \Sigma_d^4) \llbracket P \rrbracket_w \right)(\bk)
\end{align}
which are almost identical to the expressions in Section~\ref{ssec:ir_resum_pv} but with factors of $(1+f) \rightarrow 1$, as can be expected because that is the boost factor between real and redshift-space line-of-sight displacements in linear theory. The lack of these factors somewhat reduces the effect of these corrections, though they remain large on perturbative scales, and with an isotropic damping exponent instead of $\Sigma^2(\mu)$. It is worth emphasizing that these real-space velocity moments sum to the redshift space power spectrum when combined, and indeed the different factors in front of each connected component $\llbracket \ldots \rrbracket$ can be simply derived from the Leibniz rule
\begin{equation}
    (i k \mu)^n \tilde{\Xi}^{(n)}_w(\bk) = \sum_{m=0}^n {n \choose m} \left( \partial_\lambda^{n-m} e^{-\frac12 k^2 \Sigma^2(\mu)} \right)_{\lambda = 0} \llbracket \tilde{\Xi}^{(m)} \rrbracket_w(\bk)
\end{equation}
such that summing up the above contribution yields as expected $P_{s,w}(\bk) = e^{-\frac12 k^2 \Sigma^2(\mu)} (\llbracket P \rrbracket_w + (ik\mu) \llbracket v \rrbracket_w + \frac12 (ik\mu)^2 \llbracket \sigma^2 \rrbracket_w + \ldots)$.

\section{Expression for the Galaxy Density-Momentum Cross Spectrum via \texorpdfstring{$P_{LL'}^r$}{Pnmr}}
\label{app:okumura}

Using 
\begin{equation}
    \Xi^{(n)}(\bk) = \sum_{m=0}^n \binom{n}{m} (-1)^{n-m} P_{m,n-m}^r(\bk).
\end{equation}
and noting that $P_{01}^s = - \frac12 v_s$ we can rewrite Equation~\ref{eqn:mome_vk} in terms of real-space peculiar velocity spectra $P_{LL'}^r$ as 
\begin{align}
    -(ik\mu) P_{01}^s(\bk) &= -(ik\mu) P_{01}^r(\bk) + (ik\mu)^2 \left(P_{02}^r(\bk) - P_{11}^r(\bk)\right) \nonumber \\
    &+ (ik\mu)^3 \left(-\frac12 P_{03}^r(\bk) + \frac32 P_{12}^r(\bk)\right) \nonumber \\
    &+ (ik\mu)^4 \left(\frac16 P_{04}^r(\bk) - \frac23 P_{13}^r(\bk) + \frac12 P_{22}^r \right).
\end{align}
This is slightly different from Equations 2.19 and 4.1 in ref.~\cite{Okumura14} owing to a small typo in the former equation, which we re-derive from Equation 2.18 in that work to be
\begin{align}
    (-ik\mu) P_{01}^s(\bk) &= \sum_{L=1}^\infty \frac{(ik\mu)^L}{L!} (-1)^L L P_{0L}^r(\bk) + \sum_{L=1}^\infty \frac{(ik\mu)^{2L}}{(L!)^2} (-1)^L L P_{LL}^r(\bk) \nonumber \\
    &\quad \quad \quad + \sum_{L=1}^\infty \sum_{L' > L} \frac{(ik\mu)^{L+L'}}{L! L'!} (-1)^{L'} (L + L') P_{L L'}^r(\bk).
\end{align}
Noting the asymmetry in $L,L'$ in the final line and including contributions up to $L + L' = 4$ recovers the pairwise-velocity-based expression above.

\section{Gaussian Covariance}
\label{app:gaussian_covariance}

\begin{figure}
    \centering
    \includegraphics[width=\textwidth]{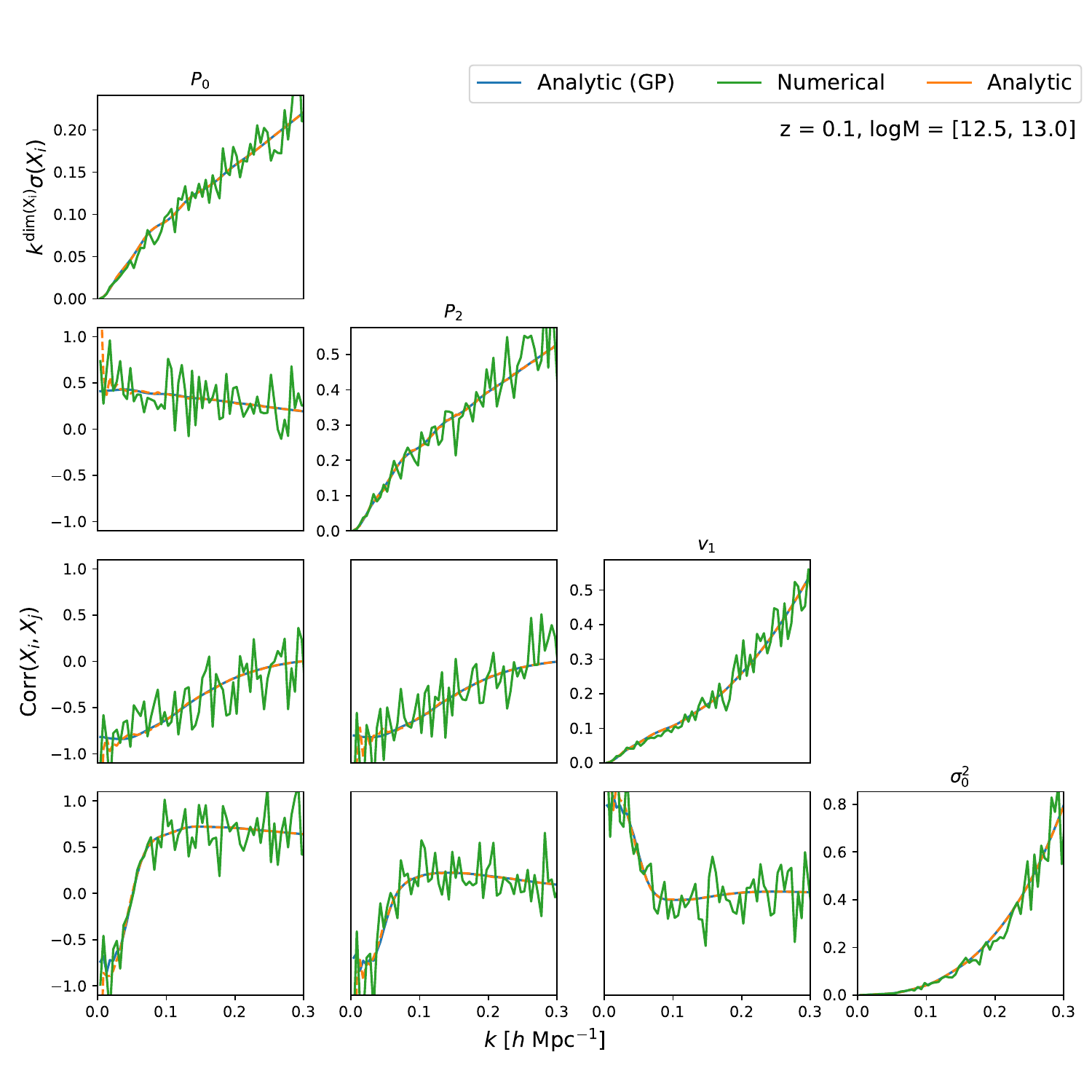}
    \caption{The covariance matrix of the redshift-space pairwise velocity spectra for halos with mass $\log M \in [12.5, 13.0]$ at $z = 0.1$, shown via its derived variances (top panels) and correlation coefficients (lower triangular panels). Analytic predictions for the disconnected component of the covariance are shown in blue (with Gaussian process smoothing) and orange (without), and agree very well with the numerical estimate from the 25 AbacusSummit boxes (green). }
    \label{fig:covariance_matrix}
\end{figure}

The estimator for velocity spectra $P_{nm}^\ell$ in a periodic box is
\begin{equation}
    \hat{P}_{nm}^\ell(k) = \frac{(2\ell + 1)}{N_k} \int \frac{d^3\bk}{(2\pi)^3} \ \hat{\rho}_n(\bk) \hat{\rho}_m(-\bk)\ \mathcal{L}_\ell(\mu),
\end{equation}
where $\rho_n = u^n (1 + \delta)$ and $N_k = V_{\rm obs} \int_\bk$ is the number of modes in bin $k$. This quantity is imaginary when $n+m$ is odd, and real when it is even, and we will in general be interested in the covariance of the real-valued imaginary and real parts rather than the complex-valued observable itself. For this reason we will also slightly abuse notation and refer to the second cumulant (pseudo-covariance) as the variance, i.e. we will avoid complex conjugates. 

We can now compute disconnected contributions to the covariance of this estimator: 
\begin{align}
    \text{cov} \{  &\hat{P}_{nm}^\ell(k),   \hat{P}_{n'm'}^{\ell'}(k) \} \nonumber \\
    &= \frac{(2\ell + 1)(2\ell' + 1)}{N_k^2} \int_{\bk,\bk'} \text{cov} \{  \hat{\rho}_n(\bk) \hat{\rho}_m(-\bk),\hat{\rho}_{n'}(\bk') \hat{\rho}_{m'}(-\bk') \} \mathcal{L}_\ell(\mu) \mathcal{L}_{\ell'}(\mu') \nonumber \\
    &= \frac{(2\ell + 1)(2\ell' + 1)}{N_k^2} \int_{\bk,\bk'} \Big( P_{nn'}(\bk) P_{mm'}(-\bk) ( (2\pi)^3 \delta_D(\bk+\bk') )^2 \nonumber \\
    & \qquad \qquad \qquad \qquad \qquad \qquad + P_{nm'}(\bk) P_{mn'}(-\bk) ( (2\pi)^3 \delta_D(\bk-\bk') )^2 \Big) \mathcal{L}_\ell(\mu)\ \mathcal{L}_{\ell'}(\mu') \nonumber \\
    &= \frac{(2\ell + 1)(2\ell' + 1)}{2 N_k} \int d\mu\ \Big( P_{nn'}(\bk) P_{mm'}(-\bk) (-1)^{\ell'} + P_{nm'}(\bk) P_{mn'}(-\bk) \Big)\ \mathcal{L}_\ell(\mu)\ \mathcal{L}_{\ell'}(\mu)
    \label{eqn:pnm_cov}
\end{align}
where we have used that $\mathcal{L}_\ell(-\mu) = (-1)^\ell \mathcal{L}_\ell(\mu)$ and $(2\pi)^3 \delta_D(\textbf{0}) = V_{\rm obs}$ and assumed azimuthal symmetry in the final line.

We can now compute the covariance of the pairwise velocity moments. Rather than writing out the full expression as in the above we will use that each element of the covariance matrix can be written as
\begin{equation}
    \text{cov} \{  \hat{\Xi}^{(n)}_\ell(k), \hat{\Xi}^{(m)}_{\ell'}(k) \} = \frac{(2\ell + 1)(2\ell' + 1)}{2 N_k} \int d\mu\ F_{nm}(\bk) \ \mathcal{L}_\ell(\mu)\ \mathcal{L}_{\ell'}(\mu)
\end{equation}
and derive the form of each $F_{nm}$. For zeroth moment we have simply $\Xi^{(0)} = P_{00}$ such that, as expected,
\begin{equation}
    F_{00}(\bk) = 2\ P_{00}(\bk)^2.
\end{equation}
Here we have used also that $P_{nm}(-\bk) = (-1)^{n+m} P_{nm}(\bk)$. The remaining components of the covariance matrix can be similarly computed to be
\begin{align*}
    F_{01}(\bk) &= 4i \ P_{00}(\bk) \text{Im}P_{01}(\bk) \nonumber \\
    F_{11}(\bk) &= 4 i^2 \left( P_{00}(\bk) P_{11}(\bk) + (\text{Im}P_{01}(\bk))^2 \right) \\
    F_{02}(\bk) &= 4\ (P_{00}(\bk) P_{02}(\bk) - (\text{Im} P_{01}(\bk))^2), \\
    F_{12}(\bk) &= 4i\ (P_{02}(\bk) \text{Im} P_{01}(\bk) - P_{00}(\bk) \text{Im} P_{12}(\bk) - 2 P_{11}(\bk) \text{Im} P_{01}(\bk) ), \\
    F_{22}(\bk) &= 4\ (P_{00}(\bk) P_{22}(\bk) + P_{02}(\bk)^2 + 2 P_{11}(\bk)^2 + 4 (\text{Im} P_{01}(\bk))(\text{Im} P_{12}(\bk))).
\end{align*}
where we have separated out imaginary spectra for ease of interpretation.

To compute the disconnected covariance matrix for our measurements from the AbacusSummit simulations suite, we measure the multipole spectra $P_{nm}$ from the same set of simulations and plug them into the above formulae. Specifically, we measure the monopole and quadrupole of even $P_{nm}$ and dipole and octopole of odd $P_{nm}$, along with the quadrupole of $P_{00}$ since it is the only higher-order multipole that has a linear-order contribution in the linear power spectrum. We then normalize these spectra by their linear-theory predictions\footnote{For spectra with $n + m > 2$ whose leading-order contributions are beyond linear order, we utilize an ansatz where velocities are contracted to form FoG-type terms, e.g. $P_{12} \sim \sigma^2 P_{01}$, and fit the resulting free coefficients roughly by eye.}, fitting the resulting fractional deviation away from linear theory $\hat{P}_{nm}^\ell / P_{nm, \rm lin}^\ell - 1$ via Gaussian process regression using a radial basis function kernel and assuming error bars on each $P_{nm}$ measurement estimated from the 25 AbacusSummit boxes. In order to avoid underestimating the modeling error due to discreteness effects on these spectra, we artificially inflate the statistical errors going into the regression by a factor of $5$.

Figure~\ref{fig:covariance_matrix} shows the variances and correlation coefficients of the redshift-space pairwise velocity spectra for halos with mass $\log M \in [12.5, 13.0]$ at $z = 0.1$. While the numerical estimate of the covariance matrix using only $25$ AbacusSummit boxes is quite noisy, it is clear that the diagonal (equal $k$) elements of the covariance matrix are very well captured by the disconnected calculations above. As a sanity check, we also include the disconnected calculation without using Gaussian process regression to smooth the measured $P_{nm}$; this makes essentially no difference at higher $k$ where the statistical uncertainities on the measurements are small. Finally, we note that the correlation coefficients between a number of observables approaches unity as $k \rightarrow 0$. This reflects that peculiar velocities have vanishing contributions from noise at low $k$ where they directly trace the large-scale gravitational potential. This makes the low-$k$ block of the covariance very sensitive to numerical errors or additional un-modeled contributions. Rather than delve further into these issues, in this work we instead opt to drop all data points below $k = 0.04h$ Mpc$^{-1}$ ($z = 0.1$) and $k = 0.035h$ Mpc$^{-1}$ ($z = 1.1$), below which the eigenvalues of the disconnected correlation matrix show numerical instabilities visible by eye.

\section{Details for Fits with Free Growth Rates}
\label{app:best_fit_curves}

\begin{figure}
    \centering
    \includegraphics[width=\textwidth]{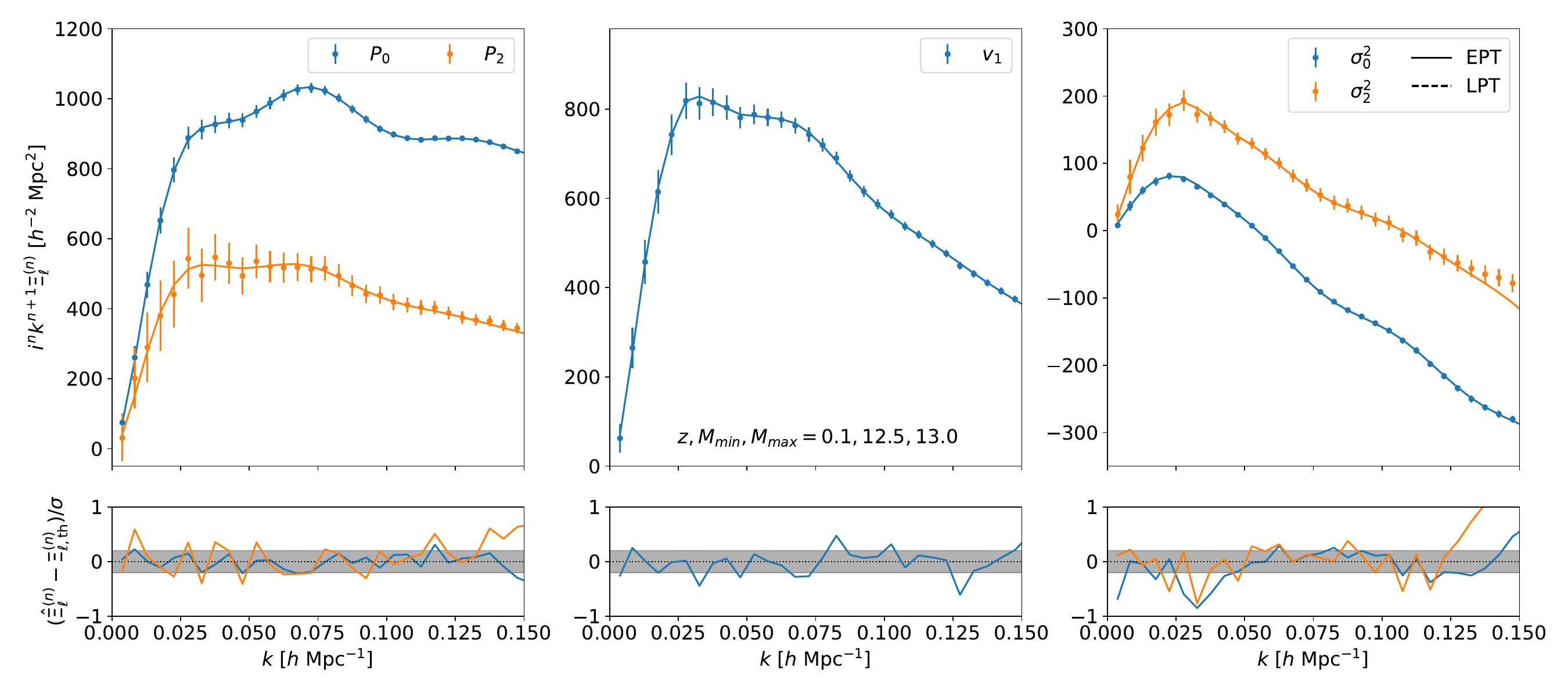}
    \includegraphics[width=\textwidth]{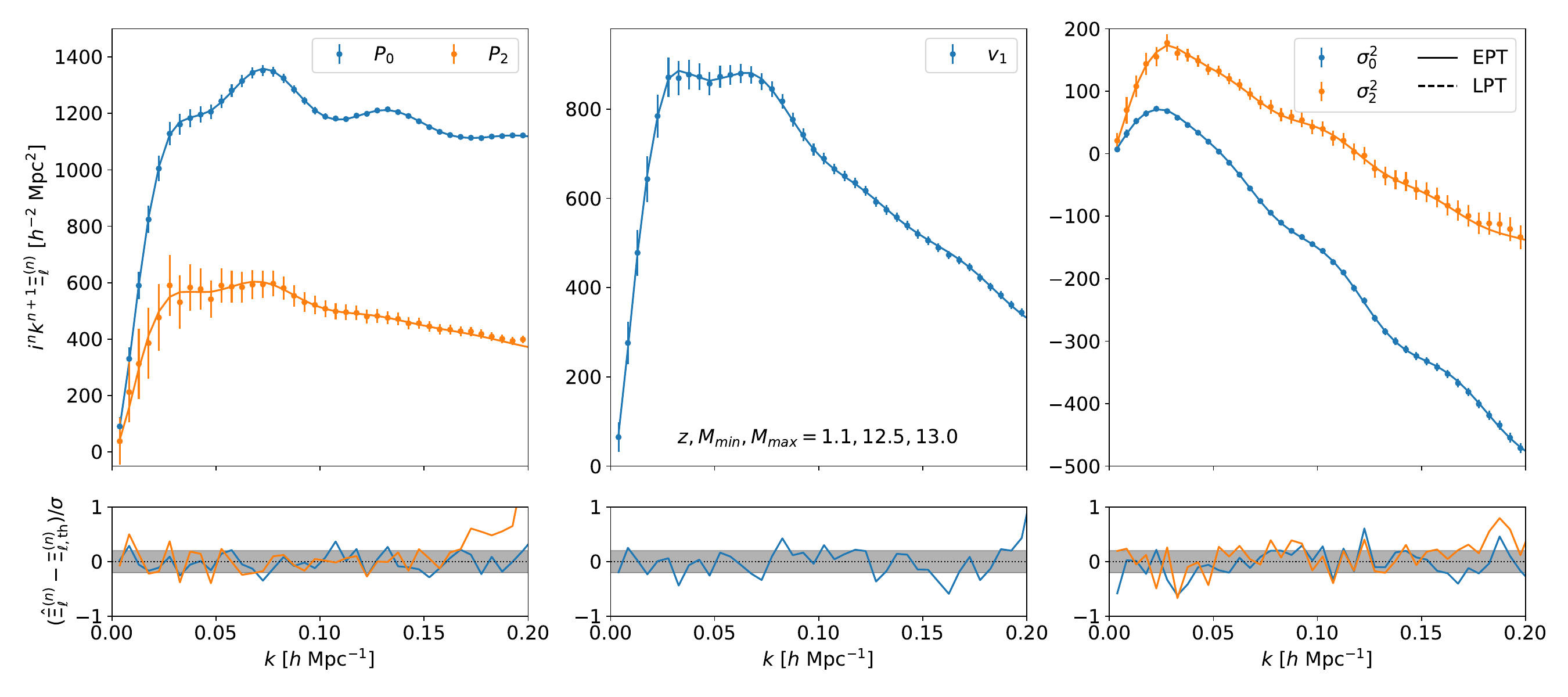}
    \caption{Best-fit curves from the cosmology-dependent fits in Section~\ref{sec:nbody}. Note that, unlike Figure~\ref{fig:fit_logM_12.5_13.0}, $1\sigma$ here corresponds to the statistical error of one AbacusSummit box; the shaded region denotes the $0.2\sigma$ region, corresponding to $1\sigma$ errors for the mean of $25$ boxes. }
    \label{fig:fz_bf}
\end{figure}

We conduct our fits using a Gaussian likelihood implemented with \texttt{Cobaya} \cite{Cobaya} with covariances as determined in Appendix~\ref{app:gaussian_covariance}. We analytically marginalize all linear parameters \cite{Taylor10}.
We use \texttt{Cobaya}'s in-built Metropolis-Hastings sampler, considering chains converged when $R - 1 < 0.05$. The best-fit curves from our fits are shown in Figure~\ref{fig:fz_bf}. We make our likelihood publicly available as part of \texttt{velocisaurus}.

\bibliography{biblio}
\bibliographystyle{jhep}

\end{document}